\documentclass[preprint,
12pt,showpacs,preprintnumbers,nofootinbib,nobibnotes,amsmath,amssymb,aps,prd]{revtex4-1}
\usepackage{epsfig}
\usepackage{graphicx}
\usepackage{bm}
\usepackage{times}
\usepackage{braket}
\usepackage{color}
\usepackage{slashed}
\usepackage{hyperref}
\usepackage{ulem}
\definecolor{Red}{rgb}{1.,0.,0.}

\definecolor{Blue}{rgb}{0.,0.,1.}

\definecolor{Green}{rgb}{0.,1.,0.}

\definecolor{Gray}{rgb}{0.5,0.5,0.5}

\definecolor{nicered}{rgb}{0.7,0.1,0.1}
\definecolor{nicegreen}{rgb}{0.1,0.5,0.1}
\hypersetup{colorlinks,citecolor=nicegreen,linkcolor=nicered}

\begin{document}

\newcommand{\beq}{\begin{eqnarray}}
\newcommand{\eeq}{\end{eqnarray}}
\newcommand{\ben}{\begin{enumerate}}
\newcommand{\een}{\end{enumerate}}
\newcommand{\non}{\nonumber\\ }
\newcommand{\jpsi}{J/\Psi}
\newcommand{\ppa}{\phi_\pi^{\rm A}}
\newcommand{\ppp}{\phi_\pi^{\rm P}}
\newcommand{\ppt}{\phi_\pi^{\rm T}}
\newcommand{\ov}{ \overline }
\newcommand{\zerot}{ {\textbf 0_{\rm T}} }
\newcommand{\kt}{k_{\rm T} }
\newcommand{\fb}{f_{\rm B} }
\newcommand{\fk}{f_{\rm K} }
\newcommand{\rk}{r_{\rm K} }
\newcommand{\mb}{m_{\rm B} }
\newcommand{\mw}{m_{\rm W} }
\newcommand{\im}{{\rm Im} }
\newcommand{\kks}{K^{(*)}}
\newcommand{\acp}{{\cal A}_{\rm CP}}
\newcommand{\pb}{\phi_{\rm B}}
\newcommand{\xeba}{\bar{x}_2}
\newcommand{\xsba}{\bar{x}_3}
\newcommand{\peas}{\phi^A}
\newcommand{\Dsl}{ D \hspace{-2truemm}/ }
\newcommand{\pvsl}{ p \hspace{-2.0truemm}/_{K^*} }
\newcommand{\esl}{ \epsilon \hspace{-1.8truemm}/ }
\newcommand{\psl}{ p \hspace{-2truemm}/ }
\newcommand{\ksl}{ k \hspace{-2.2truemm}/ }
\newcommand{\lsl}{ l \hspace{-2.2truemm}/ }
\newcommand{\nsl}{ n \hspace{-2.2truemm}/ }
\newcommand{\vsl}{ v \hspace{-2.2truemm}/ }
\newcommand{\zsl}{ z \hspace{-2.2truemm}/ }
\newcommand{\epsl}{\epsilon \hspace{-1.8truemm}/\,  }
\newcommand{\bfkk}{{\bf k} }
\newcommand{\calm}{ {\cal M} }
\newcommand{\calh}{ {\cal H} }
\newcommand{\calo}{ {\cal O} }

\def \appb{{\bf Acta. Phys. Polon. B }  }
\def \cpc{ {\bf Chin. Phys. C } }
\def \ctp{ {\bf Commun. Theor. Phys. } }
\def \epjc{{\bf Eur. Phys. J. C} }
\def \ijmpcs{{\bf Int. J. Mod. Phys. Conf. Ser.} }
\def \jhep{{\bf J. High Energy Phys. } }
\def \jpg{ {\bf J. Phys. G} }
\def \mpla{{\bf Mod. Phys. Lett. A } }
\def \npb{ {\bf Nucl. Phys. B} }
\def \plb{ {\bf Phys. Lett. B} }
\def \ppn{ {\bf Phys. Part. Nucl. } }
\def \ppnp{{\bf Prog.Part. Nucl. Phys.  } }
\def \pr{  {\bf Phys. Rep.} }
\def \prc{ {\bf Phys. Rev. C }}
\def \prd{ {\bf Phys. Rev. D} }
\def \prl{ {\bf Phys. Rev. Lett.}  }
\def \ptp{ {\bf Prog. Theor. Phys. }}
\def \zpc{ {\bf Z. Phys. C}  }
\def \jpg{ {\bf J.Phys.-G-}  }
\def \ap{ {\bf Ann. of Phys}  }

{\footnotesize


\title{The role of $D_{(s)}^\ast$ and their contributions in $B_{(s)}\to D_{(s)} h h^\prime$ decays}

\author{Jian Chai$^{1}$}\email{physchai@hnu.edu.cn}
\author{Shan Cheng$^{1,2}$}\email{scheng@hnu.edu.cn}
\author{Wen-Fei Wang$^{3,4}$}\email{wfwang@sxu.edu.cn}

\affiliation{$^1$School of Physics and Electronics, Hunan University, 410082 Changsha, China}
\affiliation{$^2$School for Theoretical Physics, Hunan University, 410082 Changsha, China}
\affiliation{$^3$Institute of Theoretical Physics, Shanxi University, 030006 Taiyuan, China}
\affiliation{$^4$State Key Laboratory of Quantum Optics and Quantum Optics Devices, Shanxi University, 030006 Taiyuan, China}

\vspace{4mm}

\date{\today}

\begin{abstract}

We demonstrate the roles of $D_{(s)}^\ast$ and their contributions in the quasi-two-body decays $B_{(s)} \to D_{(s)} h h^\prime$ 
($h, h^\prime = \{\pi, K\}$) in the perturbative QCD approach, stemming from the quark flavour changing 
$\bar{b} \to \bar{c} \, q_2 \, \bar{q}_1$ and $\bar{b} \to c \, \bar{q_1} \, \bar{q_2}$ with $q_1, q_2 = \{s/d, u\}$. 
The main motivation of this study is the measurements of significant derivations from the 
simple phase-space model in the channels $B_{(s)} \to D_{(s)} h h^\prime$ at $B$ factories and LHC, 
which is now clarified as the Breit-Wigner-tail effects from the corresponding intermediate resonant states $D_{(s)}^\ast$. 
We confirm that these effect from $D^\ast$ is small ($\sim 5\%$) in the quasi-two-body $B_{(s)} \to D \pi \pi (K)$ decaying channels, 
and predict the tiny ($< 1 \%$) contributions from $D^\ast$ in the $B_{(s)} \to D_s K \pi (K)$ decaying channels, 
our result for the $B_s \to D K \pi (K)$ decaying channels contributed only from the Breit-Wigner-tail effect of $D_s^\ast$ 
is in agreement with the current LHCb measurement. 
We recommend the Belle-II and the LHCb collaborations to restudy the processes $B^+\to \bar D^{\ast 0}\pi^+(K^+) \to D^-\pi^+\pi^+(K^+)$ 
to reveal the structure of $D^{\ast 0}$ and the strong decay $D^{\ast 0} \to D^+ \pi^-$. 

\end{abstract}

\maketitle

\section{Introduction}

Three-body $B$ decays have much richer phenomenology with the number of decaying channels being
about ten times larger than the number in two-body decays.
It provides another wonderful site to study the hadron spectroscopy and the intermediate resonant structures
with the non-trivial kinematics described by two invariant masses of three-body final states. 
From the view of QCD, it is also important to investigate the non-resonance contribution in the factorisation theorem \cite{ptep-123C01}.
In 2013, the LHCb collaboration observed the appreciable local $CP$ violation in the dalitz plot of
$B^{\pm}\to K^{\pm}\pi^+\pi^-$ and $B^{\pm}\to K^{\pm}K^+K^-$ decays \cite{prl111-101801},
which switched on a new era to study the mater-antimatter asymmetry.
In order to understand the physical observables in the full dalitz plot with abundant phase space and complicated dynamics,
the QCD-based approaches, such as the perturbative QCD (pQCD) approach \cite{plb504-6,prd63-054008,prd63-074009,ppnp51-85}
and QCD factorization (QCDF) approach \cite{prl83-1914,npb591-313,npb606-245,npb675-333}
did some pioneered studies on the quasi-two-body $B$ decays \cite{Chen:2002th,Wang:2014ira,Wang:2015uea,Ma:2016csn,
npb899-247,jhep2006-073,Huber:2020pqb,prd88-114014,prd89-074025,prd102-053006,
Zhang:2013oqa,epjc75-536,plb669-102,ElBennich:2009da}.
Furthermore, some phenomenological analyses are also implemented within the
$U$-spin, isospin and flavour SU(3) symmetries for the relevant three-body $B$ decays \cite{prd72-094031,plb728-579,
prd72-075013,prd84-056002,plb726-337,prd89-074043}.

In the traditional framework of QCD-based approaches, $D_{(s)}^\ast$ is usually treated as a stable vector meson state 
by embodying the heavy quark effective theory (HQET) \cite{Li:1999kna,Kurimoto:2002sb}.
There are two categories for the single charm two-body $B$ decays $B_{(s)} \to D_{(s)}^\ast h^\prime$, 
one is the Cabibbo-Kobayashi-Maskawa (CKM) favoured transition induced by the $b \to c$ decay \cite{Wu:1995gb,Wu:1996he},
and the other one is stemmed by the CKM suppressed $b \to u$ transition \cite{Cheng:1994fr,Keum:2003js}. 
With the interplay between $b \to c$ and $b \to u$ transitions at tree level,
the $B_{(s)} \to D_{(s)}^{(\ast)} K^{(\ast)}$ decays give the dominant constraint to the CKM angle $\gamma$ \cite{Amhis:2019ckw}. 
Theoretical studies on this type of decays are carried out with the factorization-assisted topological-amplitude (FAT) approach \cite{prd92-094016}, 
the QCDF approach \cite{ijmpa24-5845} and also the pQCD approach \cite{Li:2003wg,prd78-014018,jpg37-015002}.
Recently, collaborations at B factories \cite{prd69-112002,prd76-012006,prd79-112004,Aubert:2007xma,Wiechczynski:2014kxh} and
LHC \cite{prd90-072003,prd91-092002,prd92-012012,prd92-032002,prd94-072001}
have performed lots of dalitz analysis of the processes $B_{(s)}\to D_{(s)} h h^\prime$ and 
shown clearly the resonant structures $D^\ast_{(s)}$ in the $D_{(s)}h$ invariant mass spectroscopy, 
which without doubt enrich our knowledge of $D_{(s)}^\ast$ and promote us to study their contributions in the corresponding three-body $B$ decays.

A new issue attracted attention recently in three-body $B$ decays is the virtual contribution of a certain resonant state, 
whose pole mass is located lower than the invariant mass threshold of the final two mesons 
and the contribution is arose from the Breit-Wigner-tail (BWT) effect.  
Within the pQCD approach, the contribution from BWT effect of $\rho(770)$ is found to be half larger than 
the contribution from the pole mass of the first excited state $\rho(1450)$ in the channel $B^{\pm} \to \rho \pi^\pm \to K^+K^- \pi^{\pm}$ \cite{prd101-111901}.
Inspired by the Belle\cite{prd69-112002}, the BaBar \cite{prd79-112004} and the LHCb \cite{prd91-092002,prd94-072001} collaborations measurements of 
$B \to D\pi h$ decays, the BWT effect from the resonant state $D^\ast$ are discussed in Ref. \cite{prd99-073010} 
and calculated in the pQCD approach with the invariant mass $m_{D\pi} > 2.1 \, \mathrm{GeV}$ \cite{plb791-342}, 
showing the indispensable role of $D^{\ast 0}(2007)$ and $D^{\ast +}(2010)$ 
and indicating $\sim 5 \%$ contribution from the BWT effect to the branching ratios.  
In the channels with resonant state $D_s^\ast$, 
some derivations from the single phase-space model have also been observed at $B$ factories in the $B \to D_{s} K \pi(K)$ decays 
\cite{Aubert:2007xma,Wiechczynski:2014kxh}, moreover, 
the dalitz plot analysis from the LHCb collaboration indicates a rather large virtual contribution 
from $D^{\ast -}_s$ in the $B^0_s \to \bar{D}^0 K^- \pi^+$ decays \cite{prd90-072003}. 
These measurements motivate a systemic study of the BWT effect from $D_{(s)}^\ast$ in the three-body $B_{(s)} \to D_{(s)} h h^\prime$ decays.

In this paper we implement the pQCD approach to calculate the branching ratios of quasi-two-body decays 
$B_{(s)} \to D_{(s)}^\ast h^\prime \to D_{(s)} h h^\prime$ with in total $46$ channels,
aiming to explore the role of different resonant states, 
especially to clarify the contributions from possible BWT effect of the ground states $D^\ast$ and $D_s^\ast$. 
We will not discuss the {\it CP} violation here since there is no contributions from penguin operators 
and hence no {\it CP} asymmetry sources in the single charmed $B$ decays. 
These type of quasi-two-body $B$ decays happen in two phases, say, 
the weak decay of $b$ quark and the subsequent strong decay from the resonant states to two stable final states.
The pQCD calculation is performed in the standard formalism of two-body $B$ decays 
with replacing the single meson wave function by the di-meson one, 
in which the strong decays are represented by means of time-like form factor and parameterized by the relativistic Breit-Wigner function. 
We will also check the quasi-two-body decays in the narrow width approximation,
with which the light-cone distribution amplitude (LCDA) of di-meson system shrinks into a Delta function at the physical pole mass, 
and the result should recover the directly two-body calculation. 

The rest of the paper is organized as follows. In Sec.~\uppercase\expandafter{\romannumeral2},
we give a brief introduction for the theoretical framework.
In Sec.~\uppercase\expandafter{\romannumeral3}, the numerical results will be showed.
Discussions and conclusions will be given in Sec.~\uppercase\expandafter{\romannumeral4}.
Decaying amplitudes and the factorization formulas in the pQCD approach are collected in appendix.

\section{Framework and the (di-)meson wave functions}

In the three-body hadronic $B$ decays, all the events of final states are restricted in the Daltz plot by considering four-momentum conservation. 
Different regions in the Dalitz plot correspond to special configurations of final particle momenta: 
(a) the three corners correspond to the configuration that one hadron is soft ($E_i \sim m_i$) and 
the other two are energetic and flying back-to-back ($E_{j,k} \sim (m_B-m_i)/2$), 
(b) the intermediate parts of edges denote the kinematics that two hadrons move ahead with collinear momenta 
and the rest one recoils back ($E_i \sim m_B/2$ and $E_{j}+E_{k} \sim m_B/2$ in the massless approximation of final mesons), 
and (c) the central region in the Dalitz plot represents cases that all three hadrons are energetic and 
move fast in the space in an approximately symmetric way ($E_{i,j,k} \sim m_B/3$). 
From the QCD side, the reliable perturbative calculation can only carried out in the invariant mass region of (b) when final mesons are all light 
due to the requirement of factorisation hypothesis (energy scale to perform the perturbative calculation), 
so the physical problems in three-body hadronic $B$ decays we can handle well so far is just for the resonant state dynamics, 
which is called as quasi-two-body $B$ decays. 
In the practice, the quasi-two-body $B$ decays is usually treated as a marriage problem, 
where the first ingredient is the weak decay described by the low energy effective hamiltonian \cite{Buchalla:1995vs}
\beq
\mathcal{H}_{eff} = \frac{G_F}{\sqrt2} \, V^\ast_{qb} \, V_{q^\prime d(s)} \left[ C_1(\mu) \, O_1(\mu) + C_2(\mu) \, O_2(\mu) \right] \,,
\label{eq:eff-hamilton}
\eeq
and the cascaded ingredient is the strong decay from the resonant state $R$ to two stable mesons 
described by matrix element $\langle M_1M_2 \vert R \rangle$, 
with the energy eigenstate of $R$ writing by means of Breit-Wigner formula or others. 
In the case of $B_{(s)}\to D_{(s)}^* h^\prime\to D_{(s)}h h^\prime$ as depicted in figure \ref{fig1}, 
$R = D_{(s)}^\ast$ and $q, q^\prime \in \{c, u, d\}$, the decay amplitude can be intuitively understood by 
\beq
\mathcal{A} \left( B_{(s)}\to R \, h^{\prime} \to D_{(s)}h \, h^{\prime} \right)
= \big\langle D_{(s)}h \big\vert R \big\rangle \, \frac{1}{\left[ m^2_R - s - i m_R \, \Gamma_R(s) \right]} \, 
\big\langle R \, h^{\prime} \big\vert \, \mathcal{H}_{eff} \, \big\vert B_{(s)} \big\rangle \,,
\label{eq:quasi-2body}
\eeq
where the two matrix elements demonstrating different interactions can be studied separately by different approaches, 
and many nonperturbative parameters are involved. 
In order to calculate the quasi-two-body decaying amplitudes in an unitive theoretical framework 
with reducing the number of nonperturbative parameters as much as possible, 
the di-meson wave function, supplementing to the single meson wave functions, 
is introduced in the pQCD approach to describe all the internal dynamics happened after the weak $b$ decay. 
The decaying amplitudes is exactly written 
as a convolution of the hard kernel $H$ with the hadron distribution amplitudes (DAs) $\phi_B, \phi_{h^\prime}$ and $\phi_{D_{(s)}h}$ 
\beq
&~&\mathcal{A} \left( B_{(s)}\to R \, h^{\prime} \to D_{(s)}h \, h^{\prime} \right)
\equiv  \big\langle \left[ D_{(s)}h \right]_R h^{\prime} \big\vert \, \mathcal{H}_{eff} \, \big\vert B_{(s)} \big\rangle \non 
&=& \phi_B(x_1,b_1, \mu) \otimes H(x_i,b_i,\mu) \otimes \phi_{Dh}(x, b, \mu) \otimes \phi_{h^{\prime}}(x_3, b_3, \mu) \,,
\label{eq:quasi-2body-pQCD}
\eeq
in which $\left[ D_{(s)}h \right]_R$ indicates the di-meson system in our interesting, 
$\mu$ is the factorization scale, $b_i$ are the conjugate distances of transversal momenta. 
We present the expressions of amplitudes ${\mathcal A}$ for the considered decaying processes in the appendix.

We use the conventional kinematics in the light-cone coordinate under the rest frame of $B$ meson for the case of quasi-two-body charmed $B$ decays   
\begin{eqnarray}
&~&p_1=\frac{m_B}{\sqrt2} \left( 1, \, 1, \, 0_{\rm T} \right) \,, \quad\quad \;\;\, k_1=\left( 0, \, x_1 \, \frac{m_B}{\sqrt2}, \, k_{1{\rm T}} \right) \,, \non
&~&p_R=\frac{m_B}{\sqrt2} \left(1, \, \zeta, \, 0_{\rm T} \right) \,, \quad\quad \;\, k_R=\left( x_R \, \frac{m_B}{\sqrt2}, \, 0, \, k_{R\rm T} \right) \,, \non
&~&p_3=\frac{m_B}{\sqrt2} \left(0, \, 1-\zeta, \, 0_{\rm T} \right) \,, \quad k_3=\left( 0, \, x_3 \, (1-\zeta) \,\frac{m_B}{\sqrt2}, \, k_{3{\rm T}} \right) \,.
\end{eqnarray}
Here $p_1$ and $k_1$ represent the momentum of $B$ meson and the light spectator quark in $B$ meson, respectively, 
with $x_1$ being the longitudinal momentum fraction. 
The momentum of resonance $D_{(s)}^{\ast}$ and pseudoscalar meson $h^\prime$ are denoted by $p_R$ and $p_3$, 
with the longitudinal momentum fractions $x_R$ and $x_3$, respectively. 
The variable $\zeta \equiv p_R^2/m^2_B = s/m_B^2$ describes the momentum transfer from $B$ meson to resonance $R$, 
which in general is a function of the invariant mass of di-meson system decayed from the resonance, 
and $\sqrt{\zeta} = m_{D^{\ast}_{(s)}}/m_B$ when the invariant mass locates on the pole mass of resonance $\sqrt{s} = m_{D^{\ast}_{(s)}}$. 

Quasi-two-body charmed $B$ decays are more complicated than the charmless decays since they involve three scales, 
the mass of $B$ meson ($m_B$), the invariant mass of $D^{\ast}_{(s)}$ resonance ($\vert p_R \vert = \sqrt{s}$), 
and the mass difference of the heavy mesons and their corresponding heavy quarks ($\bar{\Lambda} \sim m_B - m_b \sim \sqrt{s} - m_c$) 
which is of the order of the QCD scale $\Lambda_{QCD}$. 
To construct a reasonable pQCD formulism of charmed $B$ decays, the following hierachies are postulated \cite{Kurimoto:2002sb} 
\beq
m_B \gg \sqrt{s} \gg \bar{\Lambda} \,,
\label{eq:hierachy-B2D}
\eeq
in which the first hierachy guarantees the power counting analysis of the hard decaying amplitude at large recoil, 
and the second hierachy justifies the power expansion in the definition of light-cone wave functions of resonant charm mesons (di-meson system). 
In our calculation, we take into account only the leading twist wave functions of heavy meson and di-meson system, say, $B$ and $D_{(s)}h$, 
and take the light meson wave functions up to twist three level with the chiral mass $m_0^\pi$ and $m_0^K$, 
then only the powers $\mathcal{O}(m_c/m_B \equiv r_c)$ and $\mathcal{O}(m_0^\pi/m_B \equiv r_0)$, 
and the momentum transfer parameter $\zeta$ appear in the expressions of charmed two-body $B$ decaying amplitudes. 
In this way the reliability of pQCD approach can be checked at least at the leading power of ${\cal O}(r_c)$ for the charmed $B$ decays 
\cite{Kurimoto:2002sb,Wu:1995gb,Wu:1996he,Cheng:1994fr,Keum:2003js}, and in this paper we are working at this accuracy. 

\begin{figure}[tb]
\begin{center}
\includegraphics[width=1.0\textwidth]{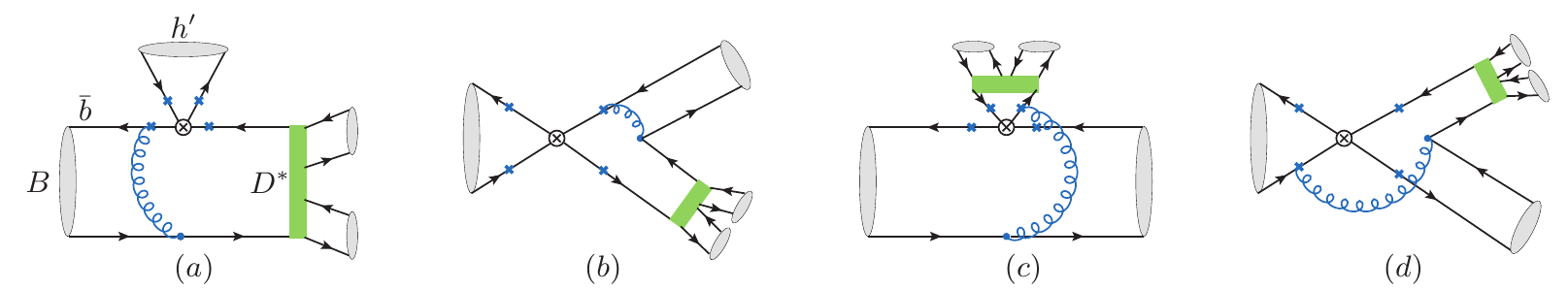}
\caption{Typical feynman diagrams for the decay processes $B_{(s)}\to D_{(s)}^\ast h^\prime \to D_{(s)} h h^\prime$, 
$h=(\pi, K)$, $h^\prime=(\pi,K)$. The symbol $\otimes$ and $\times$ denote the weak vertex and all the possible attachments of hard gluons, 
respectively, the green rectangle represents the vector states $D_{(s)}^{\ast}$.}
\label{fig1}
\end{center}
\end{figure}

We now move to the definitions of single and di-meson wave functions involved in the pQCD formulated Eq. (\ref{eq:quasi-2body-pQCD}). 
$B$ meson DAs are defined under the HQET by dynamical twist expansion \cite{Grozin:1996pq}, 
at leading twists level the nonlocal matrix element associated with $B$ meson for pQCD calculation is 
\beq
&~&\int d^4z_1 \, e^{i \bar{k}_1 \cdot z_1} 
\big\langle 0 \vert \bar{d}_\sigma(z_1) \, b_{ \beta}(0)  \vert \bar{B}^0(p_1) \big\rangle
= \frac{-i f_B}{4 N_c} \left\{ (\psl_1 + m_B) \gamma_5 
\left[ \phi_B(x_1, b_1) - \frac{\nsl_+ - \nsl_-}{\sqrt{2}} \bar{\phi}_B(x_1, b_1) \right] \right\} _{\beta\sigma} \,,
\label{eq:B-wf-2p-2}
\eeq
where the antiquark momentum aligns on the minus direction on the light cone with the momentum fraction $x_1 = k_1^-/p_1^-$. 
The transversal projection term is omitted and the integral $\varphi_{\pm} (x_1, b_1)= 
\int d k_1^+ d^2 {\bf k}_{\rm 1T} \, e^{i {\bf k}_{\rm 1T} \cdot {\bf b}_1} \, \varphi_{\pm}(k_1)$ is underlied. 
The DA $\bar{\phi}_B$ is highly suppressed by $\mathcal{O}(\ln (\bar{\Lambda}/m_B))$ in contrast to $\phi_B$ \cite{Kurimoto:2001zj}, 
we are working in the symmetry limit where $\bar{\phi}_B = 0$ to match with the current accuracy. 
The expression of DA is usually parameterized in the exponential model 
\beq
\phi_B(x_1,b_1) = N_B \,x_1^2 \, (1-x_1)^2 \, {\rm exp} \left[ - \frac{x_1^2m_B^2}{2 \omega_B^2} - \frac{(\omega_B b_1)^2}{2} \right] \,
\label{eq:B-DA}
\eeq 
with the normalization condition 
\beq
\int_0^1 dx_1 \, \phi_B(x_1, b_1=0) = 1 \,,
\label{eq:B-DA-norm}
\eeq
the first inverse moment is taken as $\omega_B = 2/3 \bar{\Lambda}$ \cite{Braun:2017liq,Beneke:2018wjp}.

Wave function of the single pseudoscalar meson $h^\prime = \pi, K$ is defined by the nonlocal matrix element \cite{Ball:2006wn,Cheng:2019ruz}, 
we here take $\pi^-$ for example,  
\beq
&~&\int d^4z_2 \, e^{i k_3 \cdot z_3} \, 
\big\langle \pi^-(p_3) \big\vert \bar{d}_\delta(z_3) \, u_\alpha(0) \big\vert 0 \big\rangle \non
&=& \frac{-i f_\pi}{4N_c} 
\, \left\{ \gamma_5 \left[ \, \psl_3 \, \phi_\pi(x_3, b_3) +  m_0^\pi \, \phi_\pi^{\rm p}(x_3, b_3) + 
m_0^\pi \left(\nsl_-\nsl_+ - 1\right) \, \phi_\pi^{\rm t}(x_3, b_3) \right] \right\}_{\alpha\delta} \,.
\label{eq:wf-P}
\eeq
Once again, the integral $\phi_\pi(x_3, b_3) = \int dk_3^+ \, d^2{\bf k_{3T}} \, e^{i {\bf k_{3T}} \cdot {\bf b_{3}}} \, \phi_\pi(k_3)$ is indicated. 
The decay constant $f_\pi$ reflects the local matrix element between the vacuum and the pion meson state, 
\beq
\big\langle \pi^+(p)  \big\vert \bar{u}(0) (\mp \gamma_\tau \gamma_5) d(0) \big\vert 0 \big\rangle = \pm i f_\pi p_{\tau} \,.
\label{eq:pi-decay-c}
\eeq 
In the expression of Eq. (\ref{eq:wf-P}), $\phi_\pi$ is the leading twist LCDA,  and $\phi_\pi^{p(t)}$ are the twist three ones. 
The light-cone vectors are defined as $n_+ = (1, 0, 0)$ and $n_- = (0, 1, 0)$, 
the chiral mass $m_0^\pi \equiv m_\pi^2/(m_u+m_d)$ originates from the equation of motion. 

Because the transversal polarized vector meson does not contribute in $B \to VP$ decays, 
we only take into account the longitudinal polarized wave function of $D_{s}h$ system. 
The hierachy $m_B \gg \sqrt{s}$ makes sure that the wave function of energetic $D_{s}h$ system absorbs the collinear dynamics, 
but with the charm quark line being eikonalized. 
This is to say, the definition of charmed meson/di-meson system wave function is the mixing of the definitions of 
$B$ meson wave function and the pion meson wave function, which is dominated by soft and collinear dynamics, respectively. 
The ${\rm P}$-wave component of $D_{(s)}h$ system with possible resonance $D_{(s)}^\ast$ is quoted as \cite{Li:1999kna,Kurimoto:2002sb}
\beq
\Phi^{\rm P}_{D_{(s)}h} = \frac{1}{\sqrt{2N_c}} \epsl_L \left( \psl_R + \sqrt s \right) \phi_{D_{(s)}h}(x,b,s).
\label{eq:wf-Dstar}
\eeq
At leading twist, the LCDA has the same Gegenbauer expansion as the vector meson $D_{(s)}^\ast$ 
\beq
\phi_{D_{(s)}h}(x,b,s)=\frac{F_{D_{(s)}h}(s) }{2\sqrt{2N_c}} \, 6 \, x \, (1-x) \left[ 1 + a_{D_{(s)}h} \, (1 - 2 \, x) \right] 
{\rm exp} \left[-\frac{\omega^2_{D_{(s)}h} \, b^2}{2} \right] \,.
\label{eq:LCDA-Dstar}
\eeq
In the ${\rm P}$-wave $D_{(s)}h$ system, the total factor reflecting the normalization 
is presented by the more general time-like form factor $F_{D_{(s)}h}(s)$, 
which in the case of single $D_{(s)}^\ast$ meson ($\sqrt{s} = m_{D_{(s)}^{\ast}}$) reduces to the decay constant $f_{D_{(s)}^\ast}$. 
In the $D_{(s)}^\ast$ dominant approximation this form factor is defined and expressed as 
\beq
F_{D_{(s)}h}(s) \equiv \frac{ s \, \bar{p}_R^\mu \, \big\langle D_{(s)} h \big\vert \bar{c} \gamma_\mu (1 - \gamma_5) q \big\vert 0 \big\rangle }
{\left[ s^2 - 2 \, s \left(m^2_{D_{(s)}} + m^2_h \right) + \left( m_{D_{(s)}}^2 - m_h^2 \right)^2 \right]} \, 
\to \frac{ \sqrt{s} \, f_R \, g_{R D_{(s)}h}}{\left[ m^2_R - s - i m_R \, \Gamma_R(s) \right]} \,,
\label{eq:ff-Dh}
\eeq
where $f_R$ is the decay constant of resonant state, 
$\bar{p}_R$ denotes the momentum difference of $D_{(s)}$ and $h$ mesons in the $D_{(s)} h$ system, 
the strong coupling is defined by means of the matrix element $g_{R D_{(s)} h} \equiv \langle D_{(s)} h \vert R \rangle$. 
We present the derivation of this approximation expression in appendix \ref{Dsast-appro}.
With the precise measurement of $g_{D^\ast D^0 \pi^+}=16.92 \pm 0.13 \pm 0.14$ \cite{prd88-052003,prl111-111801}
and the universal relation \cite{plb532-193}
\beq
\frac{g_{D^\ast D \pi^{\pm}} \, f_\pi}{2 \sqrt{m_{D^\ast} m_D}} = \frac{g_{D^\ast D_s  K} \, f_K}{2 \sqrt{m_{D^\ast} m_{D_s}}} 
= \frac{g_{D_s^\ast D K} \, f_K}{2 \sqrt{m_{D_s^\ast} m_D}}  =  g \,,
\label{eq:ghat}
\eeq
we obtain $g_{D_s^\ast D K} = 14.6 \pm0.06 \pm0.07$ and $g_{D^\ast D_s K}=14.6 \pm0.10 \pm0.13$. 
They are consistent with the result $g_{D_s^\ast D K}=14.6\pm1.7$ and $g_{D^\ast D_s K}=14.7\pm1.7$ \cite{prd74-014017} 
that are extracted from the CLEO collaboration~\cite{prd65-032003}, 
and also comparable to the predictions $g_{D_s^\ast D K} = 15.2, \, g_{D^\ast D_s K} = 15.2$ obtained from quark model \cite{jhep03-021}. 
By the way, $a_{Dh}$ is the first gegenbauer coefficient in the polynomial expansion, 
$\omega_{Dh}$ denotes the first inverse momentum of ${\rm P}$-wave $D_{(s)} h$ state, 
for these two parameters, we use the values of their partner vector meson $D^\ast_{(s)}$ in the numerical evaluation.

For the sake of generality, we go beyond the narrow resonance approximation and take the energy dependent width \cite{prd90-072003,prd91-092002,prd92-012012,prd92-032002,prd94-072001}, 
\beq
\Gamma_R(s) = \Gamma_R^{\mathrm{tot}} \left( \frac{\beta(s)}{\beta_R}\right)^3 \left(\frac{s}{m_R^2}\right) \, X^2(q r_{\rm BW}) \,,
\label{eq:width}
\eeq
in which $\Gamma_R^{\mathrm{tot}}$ is the total decay width of the resonant state. 
The non-dimensional phase space factor $\beta(s)$ of $D_{(s)} h$ system is defined by 
\beq
\beta(s) =\frac{1}{2 s} \, \sqrt{\left[s-(m_{D_{(s)}}+m_h)^2\right]\left[s-(m_{D_{(s)}}-m_h)^2\right]} \,,
\eeq
and $\beta_R \equiv \beta(m_R)$. 
In our case of the ${\rm P}$-wave configuration with $L =1$, the Blatt-Weisskopf barrier form factors $X(z)$ is \cite{barrier-ff}
\beq
X^2(q r_{\rm BW}) = \left( \frac{1 + m_R^2 \left[\beta_R  \, {\rm r_{BW}} \right]^2}{1 + s \left[ \beta(s) \, {\rm r_{BW}} \right]^2} \right) \,,
\label{eq:barrier-ff}
\eeq
the radius of the barrier is taken at ${\rm r_{BW}} = 4.0$ GeV$^{-1} \sim 0.8 \, {\rm fm}$ for all resonances \cite{Aubert:2005ce}. 
To obtain the result in Eq. (\ref{eq:barrier-ff}), we have implemented the relation $\beta(s)  \equiv q/\sqrt{s}$ 
between the phase space factor and the magnitude of momentum for the daughter meson $D_{(s)}$ or $h$. 

The total width of charged vector $D^\ast$ meson is precisely measured $\Gamma^{{\rm tot}}_{D^{\ast +}} = 83.4\pm1.8$ KeV, 
while the width of its strange partner meson $D^{\ast}_s$ is only restricted by the 
upper limit $\Gamma^{{\rm tot}}_{D^{\ast +}_s} = 1.9$ MeV so far \cite{pdg2020}. 
There are some theoretical attempts to calculate the partial width of $D_s^{\ast}$ meson, for example, 
the dominant partial width $\Gamma_{D^{\ast}_s \to D_s \gamma} = 0.066\pm0.026$ KeV is obtained in the radiative decay 
from lattice QCD evaluation \cite{prl112-212002}, 
while the prediction from QCD sum rules is about ten times larger ($\Gamma_{D^{\ast}_s \to D_s \gamma} = 0.59\pm0.15$ KeV) \cite{epjc75-243}, 
and the second dominant partial width $\Gamma_{D^{\ast}_s \to D_s \pi^0} = 8.1_{-2.6}^{+3.0}$ eV is obtained 
from the heavy meson chiral perturbation \cite{prd101-054019}. 
We will use the upper limit value in the numerical evaluation to consider the largest uncertainty\footnote{This value is also employed 
by LHCb collaboration in the study of virtual contribution from $D_s^\ast$ in the decaying channel $B^0_s\to \bar D^0 K^- \pi^+$ \cite{prd90-072003}.}. 
For the neutral vector $D$ meson, the result from the isospin analysis $\Gamma^{{\rm tot}}_{D^{\ast 0}}=55.3 \pm 1.4$ KeV 
\cite{Guo:2019qcn} consists with the value ($53$ KeV) we have extracted in the previous work \cite{plb791-342}. 
In order to access the virtual contributions (BWT effect) from the state $D_{(s)}^\ast$  
whose pole mass is lower than the threshold value of invariant mass, ie., $m_R < m_D + m_h$, 
the pole mass $m_R$ in $\beta_R$ shall be replaced by the effective mass $m^\text{eff}_0$ 
to avoid the kinematical singularity appeared in the phase space factor $\beta(s)$ \cite{prd91-092002}  
\beq
{\rm m_0^{eff}(m_0)=m^{min}+(m^{max}-m^{min}) \left[1+\tanh \left( \frac{m_0-(m^{max}+m^{min})/2}{m^{max}-m^{min}} \right) \right]}\,,
\label{eq:eff-mass}
\eeq
here ${\rm m^{max}} = m_{D_{(s)}} - m_{h^\prime}$ and ${\rm m^{min}} = m_{D_{(s)}} + m_{h} $ 
are the upper and lower thresholds of $\sqrt{s}$, respectively. 

The differential branching ratios for the quasi-two-body $B_{(s)}\to D^\ast_{(s)} h^\prime\to D_{(s)}h h^\prime$ 
decays is written as 
\beq
\frac{d{\mathcal B}}{d\zeta} = \frac{\tau_B \, q^3_{h^{\prime}} \, q^3}{48 \, \pi^3 \, m^5_B} \, \overline{\vert {\mathcal A} \vert^2} \,,
\label{eq:diff-bra}
\eeq
in which $q_{h^{\prime}}$ is the magnitude of momentum for the bachelor meson $h^{\prime}$
\beq
q_{h^{\prime}}=\frac{1}{2}\sqrt{\big[\left(m^2_{B}-m_{h^{\prime}}^2\right)^2 -2\left(m^2_{B}+m_{h^{\prime}}^2\right)s+s^2\big]/s} \,. 
\eeq

\section{Numerics and Discussions}

\vspace{-0.2cm}
\begin{table}[tb]
\begin{center}
\caption{Masses, decay constants and total widths of the mesons involved in the quasi-two-body decays.}
\label{tab1}
\vspace{4mm}
\begin{tabular}{c | c c c } 
\hline\hline
{\rm meson}  & \quad {\rm $m$(MeV)} \quad & \quad\quad {\rm $f_M$(MeV)} \quad\quad  
& $\Gamma^{\rm tot}_{D^\ast}({\rm KeV})/\tau_B(10^{-12}$ s) \\
\hline 
$\pi^\pm/\pi^0$ & 140/135 & 130 \cite{ijmpa30-1550116} & $\cdots$  \\
$K^\pm/K^0$  &  494/498  & 156 \cite{ijmpa30-1550116}& $\cdots$  \\
\hline
$D^{\ast \pm}$     & 2010 & 250 $\pm$ 11 \cite{ijmpa30-1550116}  & 83.4 $\pm$ 1.8 \\
$D^{\ast 0}$         & 2007 & 250 $\pm$ 11 \cite{ijmpa30-1550116} & 55.3 $\pm$ 1.4 \cite{Guo:2019qcn} \\
$D^{\ast \pm}_s$ & 2112 & 270 $\pm$ 19 \cite{ijmpa30-1550116} & $<$ 1900        \\
\hline
$B^{\pm}$  & 5279 & 189 \cite{prd98-074512} & 1.638 $\pm$ 0.004  \\
$B^0$        & 5280 & 189  \cite{prd98-074512} & 1.520 $\pm$ 0.004   \\
$B_s^0$    & 5367 &  231 \cite{prd98-074512} & 1.509 $\pm$ 0.004  \\
\hline\hline
\end{tabular}
\end{center}
\vspace{-0.2cm}
\end{table}

In table \ref{tab1}, we list the masses, decay constants and total widths of the mesons involved in the quasi-two-body decays. 
We take the masses and widths from PDG \cite{pdg2020}, 
use the decay constants updated from the Laplace QCD sum rules for the light and $D^\ast$ mesons \cite{ijmpa30-1550116},
and use the four-flavor lattice QCD result for the $B$ mesons decay constants \cite{prd98-074512}. 
For the first inverse moments of heavy mesons, we take $\omega_{D^{\ast 0}} = \omega_{D^{\ast \pm}} = 100 \pm 20$ MeV and 
$\omega_{D_s^{\ast 0}} = 200 \pm 40$ MeV for the vector charmed mesons, 
take $\omega_{B^{0}} = \omega_{B^\pm} = 400 \pm 40$ MeV and $\omega_{B_s^0} = 500 \pm 50$ MeV for the $B$ mesons \cite{Ali:2007ff}. 
The gegenbauer moments in the leading twist LCDAs of light mesons are taken from the QCD sum rules \cite{Ball:2006wn} as 
$a_2^\pi = a_2^K = 0.25, a_1^K = 0.06$, 
the moments of vector $D$ meson are taken from the pQCD fitting with the $B \to D^{\ast}P, D^{\ast}V$ decays data \cite{prd78-014018} as 
$a_1^{D^{\ast 0}} = a_1^{D^{\ast \pm}} = 0.5 \pm 0.1$ and $a_1^{D^{\ast 0}_s} = 0.4 \pm 0.1$. 
Besides these, the CKM matrix elements in the effective Hamiltonian 
are determined by Wolfenstein parameters $\lambda=0.22650\pm 0.00048$,  $A=0.790^{+0.017}_{-0.012}$, 
$\bar{\rho} = 0.141^{+0.016}_{-0.017}$ and $\bar{\eta}= 0.357\pm0.01$ \cite{pdg2020}, 
the masses of $D$ mesons are also taken from PDG with $m_{D^0} = 1.865$ GeV, $m_{D^\pm} = 1.870$ GeV and $m_{D_s^\pm} = 1.968$ GeV, 
the chiral masses of light mesons are chosen at $m^\pi_0 = 1.4 \pm 0.1 $ GeV and $m^K_0 = 1.9 \pm 0.1 $ GeV \cite{Ali:2007ff}, 
  
\begin{table}[tb]   
\begin{center}
\caption{The pQCD predictions of the branching ratios for quasi-two-body decays $B^0\to D_{(s)}^\ast h^\prime \to Dh h^\prime$, 
where the result of channels happened by the BWT effect are denoted by $\mathcal{B}_v$. 
Theoretical uncertainties come from the inputs of $\omega_B$, $f_{D^\ast}$, $a_{Dh}$, $A$, $\omega_{Dh}$ in turn.}
\vspace{4mm}
\label{tab2}   
\begin{tabular}{ l c c c } \hline\hline
\quad\quad\quad Decay modes & ~~~~~ ${{\mathcal B}/{\mathcal B_v}}$ & \quad Results & ~~~~~ Units\;  \\
\hline
$ B^0\to D^{\ast -} \pi^+ \to \bar D^0 \pi^- \pi^+$ &$~~~~~{\mathcal B}$
& $~~~~~~1.69^{+0.57+0.15+0.13+0.07+0.04}_{-0.52-0.15-0.11-0.05-0.02}$  &$~~~~~~10^{-3}$\; \\
$\hspace{2.4cm} \to D^- \pi^0 \pi^+$               &$~~~~~{\mathcal B}$
& $~~~~~~7.79^{+3.70+0.70+0.58+0.33+0.25}_{-2.34-0.67-0.69-0.23-0.13}$  &$~~~~~~10^{-4}$\; \\
$\hspace{2.4cm} \to D_s^-K^0 \pi^+$             &$~~~~~{\mathcal B}_v$
& $~~~~~~1.25^{+0.62+0.11+0.10+0.05+0.02}_{-0.38-0.11-0.09-0.04-0.04}$  &$~~~~~~10^{-5}$\;  \\
$B^0\to D^{\ast +}\pi^- \to D^0 \pi^+ \pi^-$         &$~~~~~{\mathcal B}$
& $~~~~~~1.01^{+0.39+0.09+0.00+0.04+0.02}_{-0.25-0.08-0.00-0.03-0.01}$   &$~~~~~~10^{-6}$\; \\
$\hspace{2.4cm} \to D^+ \pi^0 \pi^-$         &$~~~~~{\mathcal B}$
& $~~~~~~4.64^{+1.72+0.41+0.03+0.18+0.01}_{-1.22-0.40-0.00-0.13-0.02}$   &$~~~~~~10^{-7}$\; \\
$\hspace{2.4cm} \to D_s^+\bar K^0 \pi^-$         &$~~~~~{\mathcal B}_v$
& $~~~~~~1.64^{+0.59+0.15+0.00+0.06+0.00}_{-0.42-0.14-0.00-0.04-0.00}$   &$~~~~~~10^{-8}$\; \\
$B^0\to D^{\ast -}K^+ \to \bar D^0 \pi^- K^+$         &$~~~~~{\mathcal B}$
& $~~~~~~1.38^{+0.64+0.16+0.04+0.06+0.04}_{-0.42-0.09-0.10-0.04-0.03}$   &$\,~~~~~~10^{-4}$ \, \\
$\hspace{2.5cm} \to D^- \pi^0 K^+$         &$~~~~~{\mathcal B}$
& $~~~~~~6.39^{+2.78+0.57+0.56+0.27+0.05}_{-2.13-0.56-0.52-0.19-0.28}$   &$\,~~~~~~10^{-5}$ \,\\
$\hspace{2.5cm} \to D_s^-K^0 K^+$         &$~~~~~{\mathcal B}_v$
& $~~~~~~1.02^{+0.48+0.10+0.09+0.04+0.03}_{-0.31-0.09-0.08-0.03-0.02}$   &$\,~~~~~~10^{-6}$ \,\\
\hline
$ B^0\to \bar D^{\ast 0} \pi^0 \to \bar D^0 \pi^0 \pi^0$       &$~~~~~{\mathcal B}$
& $~~~~~~1.05^{+0.45+0.10+0.02+0.05+0.03}_{-0.27-0.09-0.00-0.03-0.02}$  &$~~~~~~10^{-4}$\;\\
$\hspace{2.2cm} \to D^- \pi^+ \pi^0$         &$~~~~~{\mathcal B}_v$
& $~~~~~~1.02^{+0.37+0.10+0.02+0.04+0.01}_{-0.27-0.09-0.02-0.03-0.01}$  &$~~~~~~10^{-5}$\;\\
$\hspace{2.2cm} \to D_s^- K^+ \pi^0$         &$~~~~~{\mathcal B}_v$
& $~~~~~~2.39^{+0.94+0.23+0.00+0.11+0.02}_{-0.66-0.22-0.02-0.07-0.03}$  &$~~~~~~10^{-6}$\;\\
$ B^0\to D^{\ast 0} K^0 \to D^0 \pi^0 K^0$         &$~~~~~{\mathcal B}$
& $~~~~~~9.49^{+2.57+0.85+0.40+0.38+0.19}_{-1.76-0.81-0.46-0.27-0.14}$  &$~~~~~~10^{-7}$\; \\
$\hspace{2.3cm} \, \to D^+\pi^- K^0$         &$~~~~~{\mathcal B}_v$
& $~~~~~~6.32^{+2.50+0.58+0.15+0.26+0.02}_{-1.69-0.54-0.15-0.17-0.05}$  &$~~~~~~10^{-8}$\;  \\
$\hspace{2.3cm} \, \to D_s^+K^- K^0$         &$~~~~~{\mathcal B}_v$
& $~~~~~~9.89^{+2.65+0.90+0.42+0.40+0.21}_{-2.05-0.90-0.47-0.28-0.15}$  &$~~~~~~10^{-9}$\; \\
$B^0\to \bar D^{\ast 0} K^0 \to \bar D^0 \pi^0 K^0$         &$~~~~~{\mathcal B}$
& $~~~~~~1.66^{+0.88+0.15+0.04+0.07+0.01}_{-0.45-0.14-0.05-0.03-0.02}$  &$~~~~~~10^{-5}$\; \\
$\hspace{2.3cm} \, \to D^- \pi^+ K^0$         &$~~~~~{\mathcal B}_v$
& $~~~~~~1.55^{+0.62+0.14+0.00+0.07+0.03}_{-0.43-0.13-0.00-0.05-0.03}$  &$~~~~~~10^{-6}$\; \\
$\hspace{2.3cm} \, \to D_s^- K^+ K^0$         &$~~~~~{\mathcal B}_v$
& $~~~~~~3.46^{+1.36+0.31+0.00+0.15+0.03}_{-0.94-0.30-0.00-0.10-0.03}$  &$~~~~~~10^{-7}$\; \\
\hline\hline
\end{tabular}
\end{center}
\vspace{-4mm}
\end{table}

Our predictions for totally 46 channels of quasi-two-body decays $B^0/B^+/B_s^0 \to D^\ast_{(s)} h^\prime \to D_{(s)} h h^\prime$ 
are collected in tables \ref{tab2}-\ref{tab4} in turn. 
In each table, all the possible ${\rm P}$-wave resonances $D^\ast_{(s)}$ are presented 
to clarify the strength of weak interactions in two-body $B_{(s)}$ decays, 
and the power hierarchy for the result of different two-body decays $B_{(s)} \to D^\ast_{(s)} h^\prime$ 
can be read from the weak decay amplitudes presented in appendix \ref{Amplitudes}. 
For examples, the channel $B^0 \to D^{\ast -} \pi^+$ is firstly color allowed and secondly, 
both the emission and annihilation typological diagrams give contributions. 
The channel $B^0 \to D^{\ast -} K^+$ is CKM suppressed ($\mathcal{O}(\lambda)$) and meanwhile only the emission typology contributes to the amplitude, 
while the channel $B^0 \to D^{\ast +} \pi^-$ is CKM double suppressed ($\mathcal{O}(\lambda^2)$) and simultaneously color suppressed, 
which result to the branching ratios two and three powers smaller in magnitudes than it of the channel $B^0 \to D^{\ast -} \pi^+$, respectively.
For each case of two-body decay, we go a step further to show the possible different strong couplings 
between $D^\ast_{(s)}$ and the $D_{(s)} h$ state, say, with $\bar{u}u$, $\bar{d}d$ and $\bar{s}s$ configurations. 
Once again, we take the weak decay in channel $B^0 \to D^{\ast -} \pi^+$ as the example to explain more. 
The result of the strong decays in two channels with $u$- and $d$-quark pair configurations obey 
the isospin relation $g_{D^{\ast -} \bar D^0 \pi^-} = - \sqrt{2} \, g_{\bar D^{\ast -} D^- \pi^0}$, 
of course, this relation also works for other similar two channels, 
like $D^{\ast -} \to \bar{D}^0 \pi^-$ and $D^{\ast -} \to D^- \pi^0$ which are both happened by the pole mass dynamics 
($m_{D^{\ast -}} > m_{\bar{D}^0} + m_{\pi^-}, m_{D^-} + m_{\pi^0}$). 
The strong decay $D^{\ast -} \to D_s^- K^0$ happens by the BWT effect 
and the branching ratio is apparently much smaller with comparing to the strong decays happened by the pole mass dynamics. 

\begin{table}[t]   
\begin{center}
\caption{The same as table \ref{tab2}, but for the quasi-two-body $B^+\to D_{(s)}^\ast h^\prime \to Dh h^\prime$ decaying channels.}
\label{tab3}   
\vspace{4mm}
\begin{tabular}{c c c c} \hline\hline
Decay modes          & ~~~~~${{\mathcal B}/{\mathcal B_v}}$  & \quad Results & ~~~~~ Units\;  \\
\hline
$ B^+\to D^{\ast +} \pi^0\; \to D^0 \pi^+ \pi^0$         &$~~~~~{\mathcal B}$
& $~~~~~~5.81^{+1.45+0.52+0.03+0.23+0.05}_{-1.20-0.50-0.05-0.16-0.02}$  &$~~~~~~10^{-7}$\;\\
$\hspace{2.5cm} \to D^+ \pi^0 \pi^0$         &$~~~~~{\mathcal B}$
& $~~~~~~2.65^{+0.70+0.24+0.00+0.10+0.01}_{-0.46-0.23-0.03-0.07-0.00}$  &$~~~~~~10^{-7}$\;\\
$\hspace{2.6cm} \to D_s^+\bar K^0 \pi^0$         &$~~~~~{\mathcal B}_v$
& $~~~~~~9.04^{+3.04+0.89+0.03+0.36+0.01}_{-2.26-0.79-0.01-0.25-0.01}$  &$~~~~~~10^{-9}$\;\\
\hline
$ B^+\to \bar D^{\ast 0} \pi^+\; \to \bar D^0 \pi^0 \pi^+$         &$~~~~~{\mathcal B}$
& $~~~~~~3.22^{+1.30+0.29+0.09+0.13+0.05}_{-0.94-0.28-0.22-0.09-0.03}$  &$~~~~~~10^{-3}$\;\\
$\hspace{2.6cm} \to D^- \pi^+ \pi^+$         &$~~~~~{\mathcal B}_v$
& $~~~~~~2.33^{+0.98+0.21+0.13+0.10+0.04}_{-0.72-0.20-0.16-0.07-0.05}$  &$~~~~~~10^{-4}$\; \\
$\hspace{2.7cm} \to D_s^- K^+ \pi^+$         &$~~~~~{\mathcal B}_v$
& $~~~~~~3.52^{+1.54+0.32+0.17+0.15+0.06}_{-1.03-0.30-0.17-0.10-0.07}$  &$~~~~~~10^{-5}$\; \\
$ B^+\to \bar D^{\ast 0} K^+ \to \bar D^0 \pi^0 K^+$         &$~~~~~{\mathcal B}$
& $~~~~~~2.46^{+1.18+0.14+0.16+0.10+0.09}_{-0.69-0.28-0.10-0.07-0.01}$  &$~~~~~~10^{-4}$\;\\
$\hspace{2.6cm} \to D^- \pi^+ K^+$         &$~~~~~{\mathcal B}_v$
& $~~~~~~1.80^{+0.84+0.16+0.11+0.07+0.03}_{-0.54-0.15-0.10-0.05-0.04}$  &$~~~~~~10^{-5}$\; \\
$\hspace{2.7cm} \to D_s^- K^+ K^+$         &$~~~~~{\mathcal B}_v$
& $~~~~~~2.63^{+1.18+0.24+0.13+0.11+0.04}_{-0.79-0.23-0.13-0.08-0.05}$  &$~~~~~~10^{-6}$\;  \\
$ B^+\to D^{\ast 0} K^+ \to D^0 \pi^0 K^+$         &$~~~~~{\mathcal B}$
& $~~~~~~1.00^{+0.28+0.05+0.01+0.04+0.02}_{-0.30-0.11-0.02-0.03-0.01}$  &$~~~~~~10^{-6}$\; \\
$\hspace{2.6cm} \to D^+ \pi^- K^+$         &$~~~~~{\mathcal B}_v$
& $~~~~~~5.92^{+2.23+0.53+0.12+0.24+0.05}_{-1.53-0.51-0.08-0.16-0.04}$  &$~~~~~~10^{-8}$\;  \\
$\hspace{2.7cm} \to D_s^+K^- K^+$         &$~~~~~{\mathcal B}_v$
& $~~~~~~1.10^{+0.42+0.12+0.01+0.04+0.00}_{-0.31-0.11-0.02-0.03-0.00}$  &$~~~~~~10^{-8}$\;  \\
\hline\hline
\end{tabular}
\end{center}
\end{table}

In this work, we do not take into account the pure annihilation quasi-two-body decays which is doubly suppressed by the CKM and color, 
because the pole mass contributions to their branching ratios are already rather small ($[10^{-12}, 10^{-8}]$), 
if we consider the branching ratio ($[10^{-2}, 10^{-1}]$) of the weak decay $D \to \pi K$ used to rebuilt $D$ mesons events. 
As shown in tables (\ref{tab2}-\ref{tab4}), 
the largest error in our prediction comes from the first inverse momentum of $B_{(s)}$ meson ($\omega_{B_{(s)}}$), 
the second one comes from the decay constant of intermediate resonant states ($f_{D_{(s)}^\ast}$), 
the gegenbauer moment of resonant state ($a_{Dh}$) gives the third error, 
the Wolfenstein parameter ($A$) is the fourth uncertainty source, 
and the last uncertainty source is the inverse moment of resonant state ($\omega_{Dh}$).
For the decay channels $B^+ \to D^- \pi^+ \pi^+(K^+)$ happened by the BWT effect, 
the result in this work is a litter bit larger than the previous predictions \cite{plb791-342}. 
The reason is that we here take the starting point of $D \pi$ invariant mass at their threshold value $2.01$ GeV, 
while the evaluation is chosen to start at $2.1$ GeV in the former work. 

\begin{table}[tb]   
\begin{center}
\caption{The same as table \ref{tab2}, but for the quasi-two-body $B_s^0\to D_{(s)}^\ast h^\prime \to Dh h^\prime$ decaying channels.}
\label{tab4}   
\vspace{4mm}
\begin{tabular}{ l c c c } 
\hline\hline
\quad\quad\quad  Decay modes   & ~\; ${{\mathcal B}/{\mathcal B_v}}$  & \quad Results & ~~~~~ Units\;  \\
\hline
$ B_s^0\to D_s^{\ast -} \pi^+ \to \bar D^0 K^- \pi^+$         &$~~~~~{\mathcal B}_v$
& $~~~~~~1.90^{+0.94+0.28+0.16+0.08+0.14}_{-0.59-0.26-0.14-0.06-0.15}$    &$~~~~~~10^{-5}$\; \\
$\hspace{2.4cm} \to  D^- \bar K^0 \pi^+$         &$~~~~~{\mathcal B}_v$
& $~~~~~~1.83^{+0.94+0.27+0.15+0.08+0.14}_{-0.57-0.25-0.13-0.06-0.14}$\   &$~~~~~~10^{-5}$\; \\
$\hspace{3.1cm} ---$              &$~~~~~-$    &~~~~~~$---$\; & ~~~~~~$-$\    \\
$ B_s^0\to D_s^{\ast -} K^+ \to \bar D^0 K^- K^+$         &$~~~~~{\mathcal B}_v$
& $~~~~~~1.28^{+0.66+0.19+0.10+0.06+0.10}_{-0.42-0.17-0.09-0.04-0.10}$    &$~~~~~~10^{-6}$\;\\
$\hspace{2.5cm} \to  D^- \bar K^0 K^+$         &$~~~~~{\mathcal B}_v$
& $~~~~~~1.23^{+0.66+0.18+0.09+0.05+0.09}_{-0.40-0.17-0.09-0.04-0.10}$    &$~~~~~~10^{-6}$\; \\
$\hspace{3.1cm} ---$          &$~~~~~-$         &~~~~~~$---$\; & ~~~~~~$-$\    \\
\hline
$ B_s^0\to D^{\ast -} \pi^+ \to \bar D^0 \pi^- \pi^+$        &$~~~~~{\mathcal B}$
& $~~~~~~8.61^{+0.76+0.77+0.84+0.37+0.19}_{-0.84-0.74-0.97-0.26-0.21}$    &$~~~~~~10^{-7}$\;\\
$\hspace{2.4cm} \to  D^- \pi^0 \pi^+$         &$~~~~~{\mathcal B}$
& $~~~~~~3.82^{+0.70+0.34+0.36+0.16+0.06}_{-0.23-0.33-0.29-0.12-0.08}$    &$~~~~~~10^{-7}$\; \\
$\hspace{2.4cm} \to  D_s^-K^0 \pi^+$         &$~~~~~{\mathcal B}_v$
& $~~~~~~5.50^{+0.44+0.49+0.82+0.24+0.13}_{-0.48-0.47-0.73-0.17-0.13}$    &$~~~~~~10^{-9}$\; \\
$B_s^0\to D^{\ast +} K^- \to D^0 \pi^+ K^-$         &$~~~~~{\mathcal B}$
& $~~~~~~1.13^{+0.46+0.10+0.00+0.04+0.00}_{-0.31-0.10-0.00-0.03-0.00}$    &$~~~~~~10^{-6}$\; \\
$\hspace{2.5cm} \to  D^+ \pi^0 K^-$         &$~~~~~{\mathcal B}$
& $~~~~~~5.14^{+2.10+0.46+0.01+0.20+0.00}_{-1.41-0.44-0.00-0.14-0.00}$    &$~~~~~~10^{-7}$\;\\
$\hspace{2.5cm} \to  D_s^+\bar K^0 K^-$         &$~~~~~{\mathcal B}_v$
& $~~~~~~1.67^{+0.68+0.15+0.00+0.06+0.00}_{-0.46-0.14-0.00-0.04-0.00}$    &$~~~~~~10^{-8}$\;\\
\hline
$ B_s^0\to \bar D^{\ast 0} \pi^0 \to \bar D^0 \pi^0 \pi^0$  &$~~~~~{\mathcal B}$
& $~~~~~~4.16^{+0.24+0.37+0.50+0.18+0.07}_{-0.44-0.36-0.41-0.13-0.16}$    &$~~~~~~10^{-7}$\;\\
$\hspace{2.3cm} \to  D^- \pi^+ \pi^0$         &$~~~~~{\mathcal B}_v$
& $~~~~~~2.36^{+0.25+0.21+0.28+0.10+0.05}_{-0.21-0.20-0.27-0.07-0.06}$    &$~~~~~~10^{-8}$\;\\
$\hspace{2.3cm} \to  D_s^- K^+ \pi^0$         &$~~~~~{\mathcal B}_v$ 
& $~~~~~~2.75^{+0.23+0.25+0.41+0.12+0.06}_{-0.24-0.24-0.37-0.08-0.01}$    &$~~~~~~10^{-9}$\;\\
$ B_s^0\to \bar D^{*0} \bar K^0 \to \bar D^0 \pi^0 \bar K^0$ &$~~~~~{\mathcal B}$
& $~~~~~~1.71^{+0.73+0.15+0.04+0.07+0.02}_{-0.56-0.15-0.04-0.05-0.08} $   &$~~~~~10^{-4}$ \\
$\hspace{2.3cm} \to  D^- \pi^+ \bar K^0$         &$~~~~~{\mathcal B}_v$
& $~~~~~~1.59^{+0.75+0.14+0.01+0.07+0.01}_{-0.50-0.14-0.00-0.05-0.02}$    &$~~~~~10^{-5}$ \\
$\hspace{2.3cm} \to  D_s^- K^+ \bar K^0$         &$~~~~~{\mathcal B}_v$
& $~~~~~~3.82^{+1.79+0.34+0.01+0.16+0.03}_{-1.18-0.33-0.00-0.12-0.04}$    &$~~~~~10^{-6}$\\
\hline\hline
\end{tabular}
\end{center}
\end{table}

\begin{figure}[tb]
\vspace{-4mm}
\begin{center}
\includegraphics[width=0.49\textwidth]{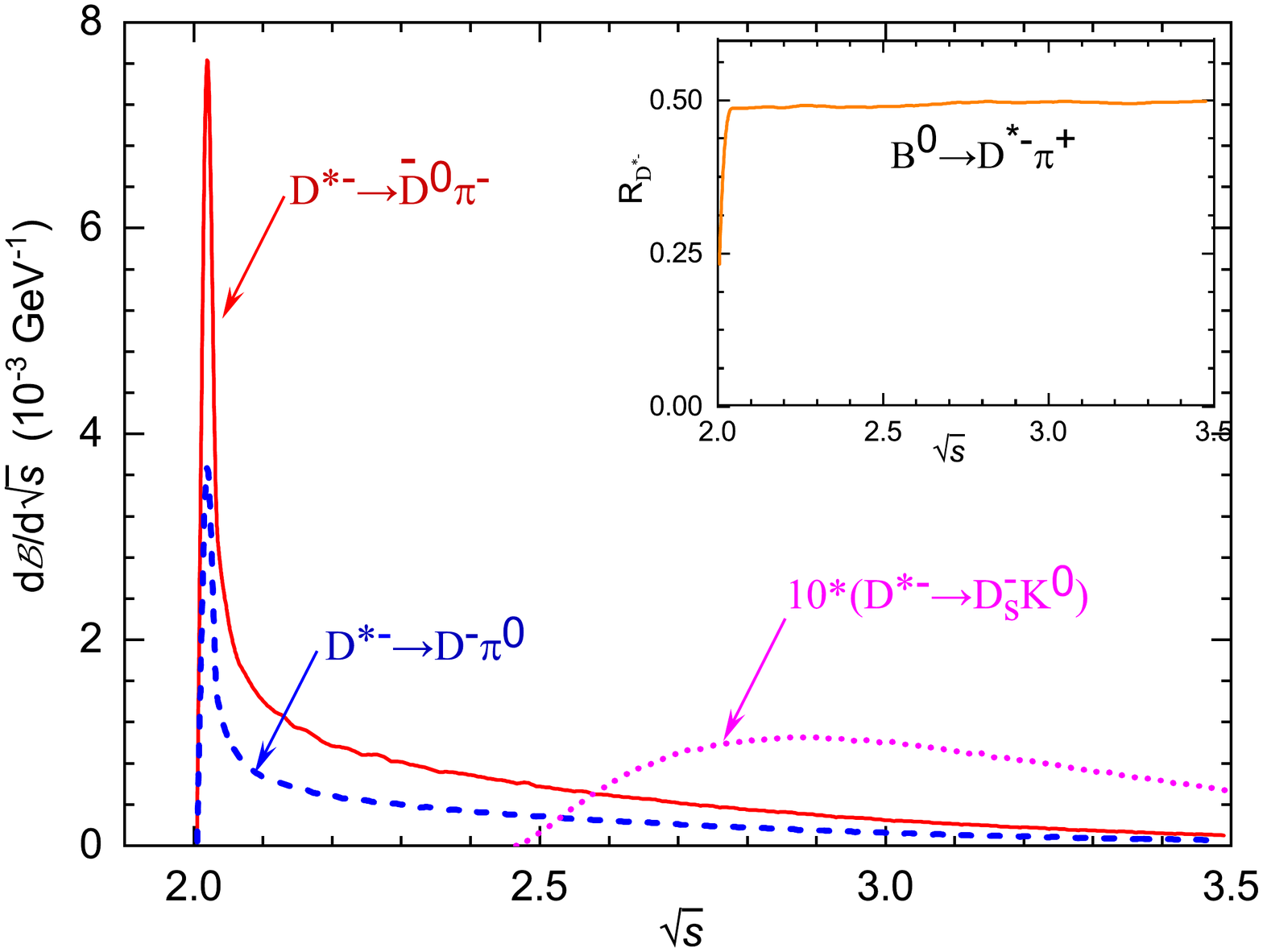}
\includegraphics[width=0.49\textwidth]{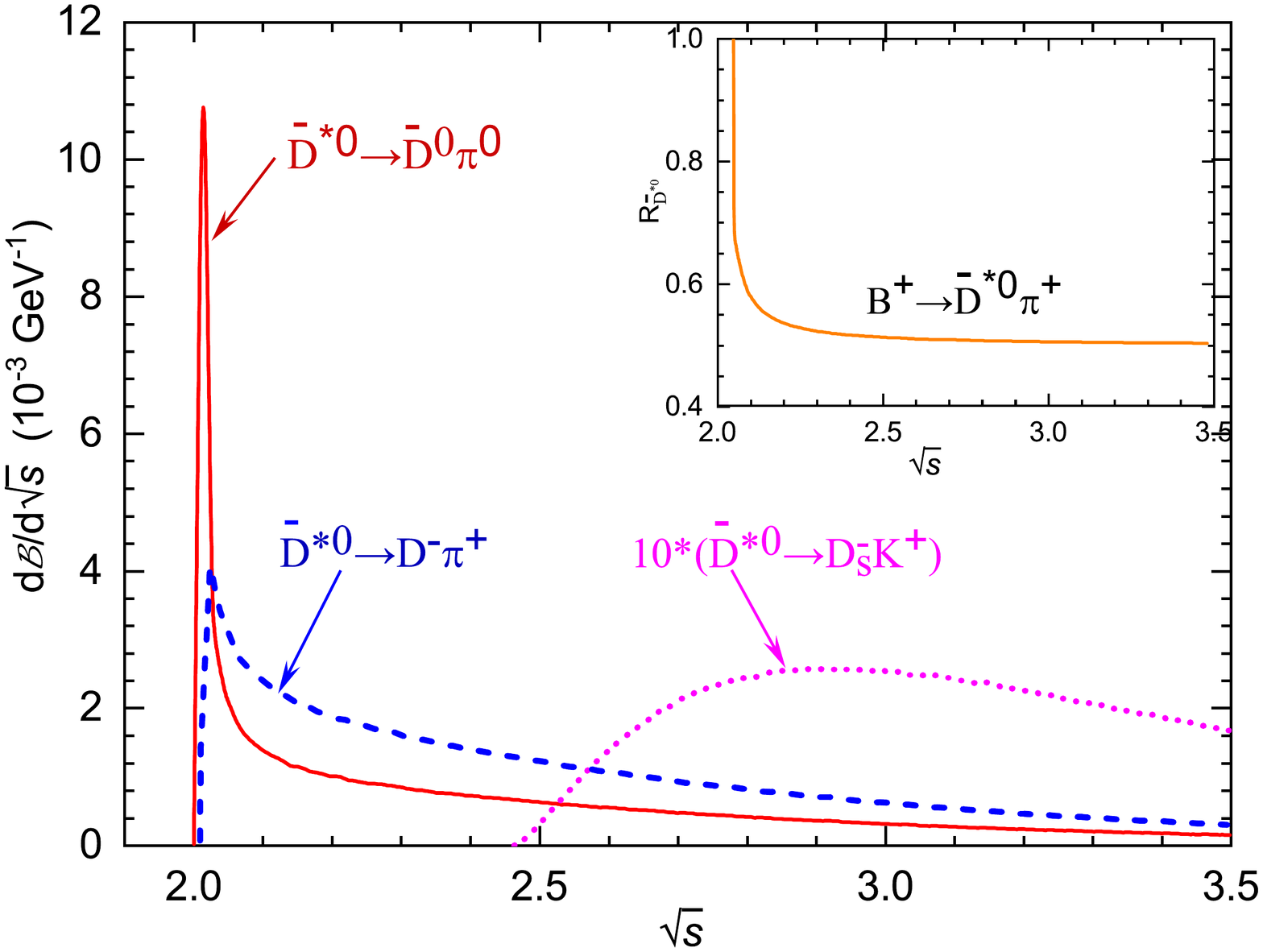}
\vspace{-6mm}
\caption{The differential branching ratios for the quasi-two-body $B^0\to D^{\ast -}\pi^+\to Dh\pi^+$ decays (left pannel) and
$B^+\to \bar D^{\ast 0}\pi^+\to Dh\pi^+$ decays (right panel). The embedded graphs denote the ratios $R_{D^{\ast} \to D \pi}$. }
\label{fig23}
\end{center}
\end{figure}

In figure \ref{fig23}, we depict the differential branching ratios of channels $B^0\to D^{\ast -}\pi^+\to Dh\pi^+$ and 
$B^+\to \bar D^{\ast 0}\pi^+\to Dh\pi^+$ to reveal the relative strength in different strong couplings following a same two-body weak decay.  
The processes $D^{\ast -} \to \bar D^0 \pi^- \, ({\rm in \, red}), \, D^- \pi^0 \, ({\rm in \, blue})$ happened by pole mass dynamics 
and the process $D^{\ast -} \to D_s^- K^0 \, ({\rm in \, magenta})$ happened by BWT effect, 
following the same $B^0\to \bar D^{\ast -}\pi^+$ weak decay, are shown explicitly on the left panel. 
In parallel, the pole mass dynamical process $\bar D^{\ast 0} \to \bar D^0 \pi^0 \, ({\rm in \, red})$ and 
the BWT induced processes $\bar D^{\ast 0} \to D^- \pi^+ \, ({\rm in \, blue}), \, D_s^- K^+ \, ({\rm in \, magenta})$, 
following the $B^+\to \bar D^{\ast 0}\pi^+$ weak decay are shown on the right panel. 
Within the embedded graphs, we display the evolution of isospin ratios on the invariant mass,  
\beq
&&R_{D^{\ast-} \to D \pi} \equiv \frac{d\mathcal{B}(B^0\to D^{\ast -}\pi^+\to D^- \pi^0 \pi^+)}
{d\mathcal{B}(B^0\to D^{\ast -}\pi^+\to \bar D^0 \pi^- \pi^+)} \,, \non
&&R_{\bar D^{\ast 0} \to D \pi} \equiv \frac{d\mathcal{B}(B^+\to \bar D^{\ast 0}\pi^+\to \bar D^0 \pi^0 \pi^+)}
{d\mathcal{B}(B^+\to \bar D^{\ast 0}\pi^+\to  D^- \pi^+ \pi^+)} \,.
\eeq
It can be seem that the ratio $R_{D^{\ast \pm}}$ goes to $0.5$ suddenly around the $D^{\ast \pm}$ pole, 
which is natural due to the pole mass dynamics for both the strong decays $D^{\ast -} \to D^- \pi^0$ and $D^{\ast -} \to \bar{D}^0 \pi^-$. 
Nevertheless, the ratio $R_{D^{\ast 0}}$ trends to $0.5$ from infinity smoothly, 
the underlying reason is that the $\bar D^{\ast 0}\to  D^- \pi^+ $ process happens by the BWT effect 
with the threshold value of $D^-\pi^+$ state being a litter bit larger than the $\bar D^{\ast 0}$ pole mass, 
hence the peak of its $d\mathcal{B}/d\sqrt{s}$ curve emerges at the invariant mass a lit bitter larger than the pole mass. 
Concerning on the BWT effect in the channels $B^0 \to D^{\ast -} \pi^+ \to D_s^- K^0 \pi^+$ and 
$B^+ \to D^{\ast 0} \pi^+ \to D_s^- K^+ \pi^+$ (${\rm in \, magenta}$) induced by the $s$-quark pair configuration, 
we multiply their result by ten to show clearly for the evolution behaviour. 
Their curves are smooth and the locations of the largest distribution ($2.8 - 2.9$ GeV) are far away from the resonance pole masses 
by $0.8 - 0.9$ GeV with the threshold values of $D_s K$ states being larger than their resonance pole masses by $0.455$ GeV.

\begin{figure}[t]
\vspace{-4mm}
\begin{center}
\includegraphics[width=0.6\textwidth]{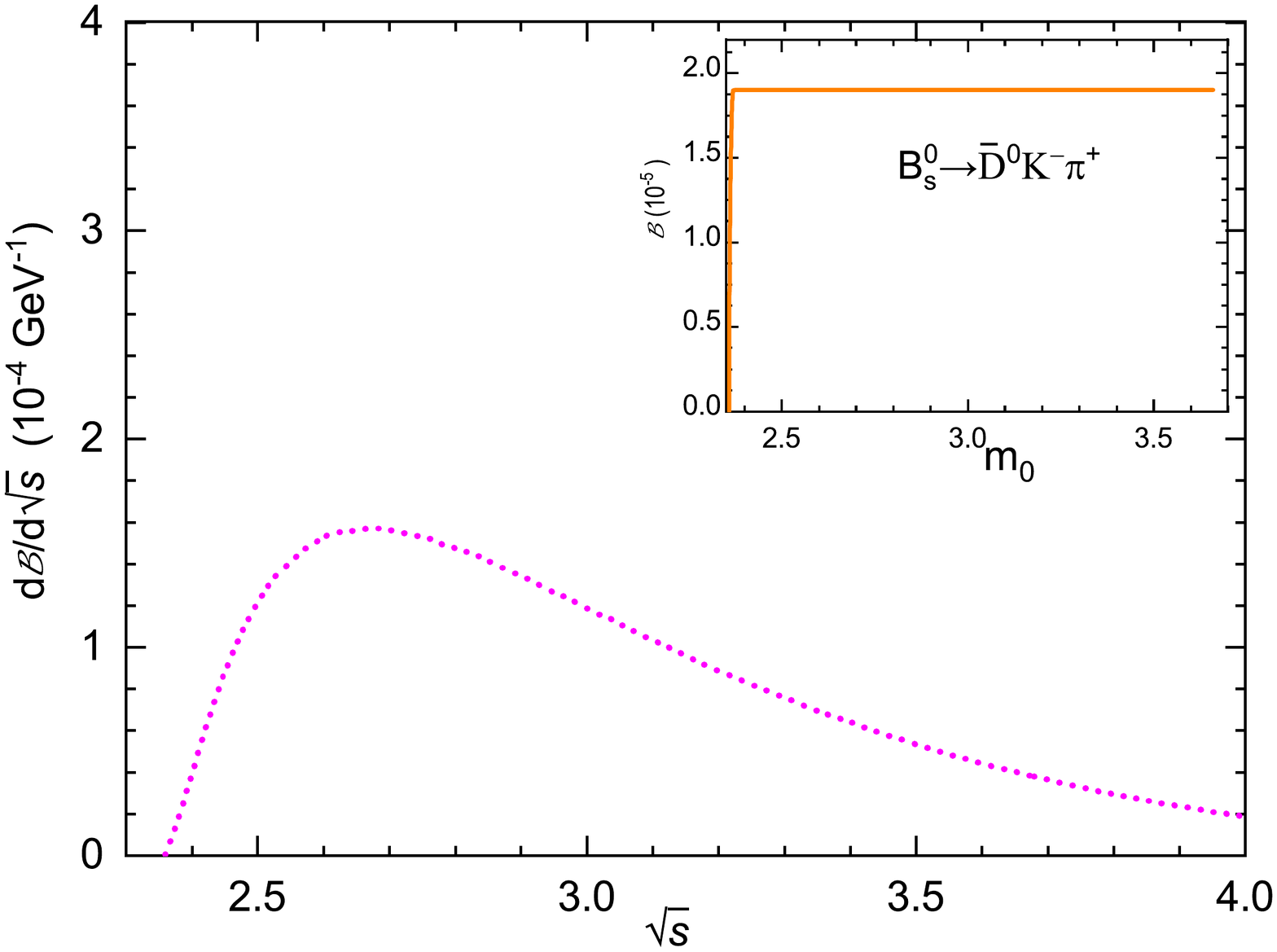}
\vspace{-6mm}
\caption{The differential branching ratios for the quasi-two-body decay $B_s^0\to D_s^{\ast -}\pi^+\to \bar D^0 K^-\pi^+$ 
with the invariant mass $\sqrt{s} \in [2.3, 4.0]$ GeV. 
The embedded graph indicates the evolution on $m_0$ from the $\bar D^0 K^-$ threshold value to $m_0^{{\rm eff}}$.}
\label{fig4}
\end{center}
\end{figure}

We plot the differential branching ratio of $B_s^0\to D_s^{\ast -}\pi^+\to \bar D^0 K^-\pi^+$ decay with 
the invariant mass of $\bar D^0 K^-$ state varying in $[2.3, 4.0]$ GeV in figure \ref{fig4}, 
we also embed the evolution of total branching ratio on $m_0$. 
\begin{itemize}
\item 
As it is shown in the embedded graph, the total branching ratio does not displace a dependence on the effective mass, 
that is to say, the width effect in Eq. (\ref{eq:width}) of $D^{\ast -}_s$ is negligible here. 
This can be understood by the Breit-Wigner expression in Eq. (\ref{eq:quasi-2body}), 
where the real part is much larger than the imaginary part in the denominator, say, 
$\vert m^2_{D^{\ast -}_s} - s \vert \gg \vert m_{D^{\ast -}_s} \, \Gamma_{D^{\ast -}_s}(s)\vert $, 
when $\sqrt{s} > 2.359$ GeV and the total width $\Gamma_{D^{\ast -}_s} < 1.9$ MeV. 
\item
One can see that the BWT effect in this channel is at the same order as in the channels of 
$B^0 \to D^{\ast -} \pi^+ \to D_s^- K^0 \pi^+$ and $B^+ \to \bar{D}^{\ast 0} \pi^+ \to D_s^- K^+ \pi^+$ (magenta curves in figure \ref{fig23}), 
even though the total decay width of $D^\ast_s$ is about twenty larger than the width of $D^\ast$. 
This is not surprise because the BWT effect is not sensitive to the width but to the real part of Breit-Wigner propagators, 
as we have demonstrated above.  
\item
For the real part of Breit-Wigner propagators in this channel, 
the threshold value $2.359$ GeV is more closer to the resonance pole mass $m_{D^{\ast -}_s} = 2.112$ GeV, 
comparing to the interval between di-meson threshold and resonance pole mass in the two $B$ meson decaying channels ($0.455$ GeV), 
so the location of the largest distribution in this channel is more closer to the $\bar D^0 K^-$ threshold. 
\item
We highlight that the BWT effect discussed in this work is not equal to the width effect of the intermediate resonances, 
it is determined by the real part of Breit-Wigner propagator since $\vert m_{R}^2 - s \vert \gg m_{R} \Gamma_{R}(s)$ in the case of $R = D^\ast_{(s)}$. 
That's why we get the significant contribution from the BWT effect for the quasi-two-body $B_{(s)} \to D^\ast_{(s)} h^\prime \to D_{(s)}h h^\prime$ decays, 
even though the total widths of $D^\ast_{(s)}$ mesons are rather small. 
\end{itemize}

We compare our predictions with the available measurements for some channels induced by the BWT effect in table \ref{tab5}, 
where the theoretical and experimental errors are both added in quadrature. 
The result listed in the second column is obtained with the integral of invariant mass starting from the threshold value. 
Because the data of the first two channels is obtained by taking the integral with a cut $\sqrt{s} \geq 2.1$ GeV which is a litter bit larger than the threshold, 
we show the pQCD predictions with the same cut in the third column, denoting by ${\mathcal B}_v^{{\rm cut}}$. 
For the channel $B^+\to \bar D^{\ast 0}\pi^+ \to D^-\pi^+\pi^+$, the prediction is more inclined to the Belle data, but still with a large uncertainty.  
For the channel $B^+\to \bar D^{\ast 0}K^+ \to D^-\pi^+K^+$, 
the central value of pQCD prediction is about two times larger in magnitude than the LHCb measurement, 
even though a large uncertainty is associated with experiment data. 
With considering the well consistence between the measurements and the pQCD predictions for the 
relevant two-body weak decays $B^+ \to \bar D^{\ast 0} \pi^+$ and $B^+ \to \bar D^{\ast 0} K^+$ as shown in table \ref{tab6}, 
we believe the result of this two channels $B^+ \to \bar D^{\ast 0} \pi^+ \to D^- \pi^+ \pi^-$ and $B^+ \to \bar D^{\ast 0} K^+  \to D^- \pi^+ K^+$
have the similar power behaviour as displayed in table \ref{tab5} because they process the same strong decay. 
We hope that the Belle-II and the LHCb collaborations restudy these two channels to reveal the important information of $D^{\ast 0}$ 
and the strong decay $D^{\ast 0} \to D^+ \pi^-$. 
For the channel $B_s^0\to D_s^{\ast -} \pi^+ \to \bar D^0 K^- \pi^+$, the pQCD prediction is in the same power as the data, 
more data will explain more. 

\begin{table}[t]
\begin{center}  
\caption{The comparison between pQCD predictions and available experimental measurements for some channels happened by the BWT effect.}
\label{tab5}
\vspace{4mm}
\begin{tabular}{c c c r} 
\hline\hline
Decay modes    & \quad\quad${\mathcal B}_v (10^{-5})$ \quad\quad  &  
\quad\quad${\mathcal B}_v^{{\rm cut}} (10^{-5})$ \cite{plb791-342} \quad\quad &   \quad\quad Data $(10^{-5})$ \quad\quad \\  
\hline
$B^+\to \bar D^{\ast 0}\pi^+ \to D^-\pi^+\pi^+$ &  $23.3^{+10.1}_{-7.70}$ &
$19.2^{+8.80}_{-6.20} $  &  $22.3\pm3.20 $ \, \cite{prd69-112002}  \\
 ~    & ~ & ~ &$ 10.9 $ \, \cite{prd79-112004} \\ 
 ~    & ~ & ~ & $ 10.9 \pm 2.70$ \, \cite{prd94-072001} \\
$B^+\to \bar D^{\ast 0}K^+\to D^-\pi^+K^+$  &$1.80^{+0.86}_{-0.57}$ & $1.48^{+0.68}_{-0.47} $ &$0.56 \pm 0.23 $ \, \cite{prd91-092002} \\
$B_s^0 \to D_s^{\ast -} \pi^+ \to \bar D^0 K^- \pi^+$  & $1.90^{+1.01}_{-0.68}$ & ~ & $4.70 \pm 4.38$ \, \cite{prd90-072003} \\
\hline\hline
\end{tabular}
\end{center}
\vspace{-0.2cm}
\end{table}

With the calculations for these quasi-two-body decays, 
we can extract the branching ratios of single charmed two-body $B$ decays by using the narrow width approximation
\beq
{\mathcal B}(B\to D_{(s)}^\ast h^{\prime} \to D_{(s)} h h^{\prime}) \approx 
{\mathcal B}(B\to D_{(s)}^\ast h^{\prime}) \cdot {\mathcal B}(D_{(s)}^\ast \to D_{(s)} h) \,,
\label{eq:narrow-appro}
\eeq
with the measurements $\mathcal{B}( D^{\ast  +}\to D^0\pi^+)=67.7\%, \, \mathcal{B}( D^{\ast +}\to D^+\pi^0)=30.7\%$ and $\mathcal{B}( D^{\ast 0}\to D^0\pi^0)=64.7\%$ \cite{pdg2020}. 
The result of the CKM favoured channels is shown in table \ref{tab6}, 
which is consistent with the direct two-body calculations and agree with the data. 
For the CKM suppressed decays, only the channel $B^+\to D^{\ast 0} K^+$ has been measured 
with the branching ratio $\mathcal{B}(B^+\to D^{\ast 0} K^+) = (7.8 \pm 2.2)\times 10^{-6}$ \cite{pdg2020}, 
our extraction here gives $(1.54^{+0.45}_{-0.49}) \times 10^{-6}$, 
consisting with the result $(0.71^{+0.76}_{-0.53}) \times 10^{-6}$ by direct two-body calculation from pQCD approach \cite{jpg37-015002}, 
however, deviating from the data by $3 \sigma$. 
The result from FAT approach $(11.8^{+3.5}_{-3.5})\times 10^{-6}$ is in agreement with the data, 
because their nonfactorizable annihilation-type contribution is fit from data and much larger than it calculated from the pQCD approach. 
LHCb will accumulate much more data to clarify this problem. 

\begin{table}[t]  
\begin{center}
\caption{Branching ratios of $B_{(s)} \to D^{\ast} h^\prime$ decays obtained from quasi-two-body processes under the narrow width approximation. 
The previous two-body pQCD calculation \cite{prd78-014018} and the experimental measurements are also listed for comparison.}
\label{tab6}   
\vspace{4mm}
\begin{tabular}{l c c c} 
\hline\hline
Decay modes \quad & \quad\quad\quad pQCD $(10^{-4})$ \quad\quad & \quad\quad This work $(10^{-4})$  \quad\quad 
& \quad\quad Data $(10^{-4})$ \cite{pdg2020} \quad\quad  \\
\hline
$B^0\to D^{\ast -} \pi^+$  &$26.1^{+8.90}_{-9.50} $ & $25.0^{+8.90}_{-8.10}$ & $27.4\pm1.30$\\
~&  ~&  $25.4^{+12.4}_{-8.20}$  & \\
$B^0\to D^{\ast -}K^+$     &$2.21^{+0.82}_{-0.83} $ & $2.04^{+0.98}_{-0.65}$ & $2.12\pm0.15$\\
&  & $2.08^{+0.94}_{-0.74}$    & \\
$B^0\to \bar D^{\ast 0} \pi^0$   &$2.30^{+0.87}_{-0.83} $ & $1.62^{+0.71}_{-0.44}$ & $2.20\pm0.60$\\
$B^0\to \bar D^{\ast  0} K^0$    &$0.25^{+0.10}_{-0.09} $ & $0.26^{+0.14}_{-0.07}$ & $0.36\pm0.12$\\
$B^+\to \bar D^{\ast 0} \pi^+$   &$51.1^{+14.7}_{-14.2} $ & $49.8^{+20.7}_{-15.6}$ & $49.0\pm1.70$\\
$B^+\to \bar D^{\ast  0} K^+$    &$3.94^{+1.24}_{-1.32} $ & $3.80^{+1.86}_{-1.16}$ & $3.97^{+0.31}_{-0.28}$\\
$B_s^0 \to \bar D^{\ast  0} \bar K^0$  &$4.14^{+2.01}_{-1.52} $ & $2.64^{+1.15}_{-0.90} $ &$2.80\pm1.10$\\
\hline\hline
\end{tabular}
\end{center}
\end{table}

\section{Conclusion}

In this paper we studied systematically the role of $D^{\ast}_{(s)}$ and their contributions 
in $B_{(s)} \to D_{(s)} h \, h^\prime$ ($h^\prime = \pi, K$) decays 
by taking the di-meson LCDAs of $D_{(s)} h$ system in the framework of the pQCD approach. 
With the weak decays $B_{(s)} \to D^{\ast}_{(s)} h^\prime$ stemming only from the tree level current-current operator, 
there are no source of weak phase differences to generate {\it CP} violations, so we predicted only the branching ratios. 
In the local kinematic region where the invariant mass of final $D_{(s)} h$ system locates in/around the interval of $D^{\ast}_{(s)}$ 
and simultaneously the other invariant mass approaching zero, 
three-body decay can be treated as a quasi-two-body decaying process and divided into two ingredients. 
The first one is the $b$ quark weak decays which have the same formula as in the two-body $B$ decays, 
and the strong decays happened subsequently are absorbed into the di-meson wave functions of $D_{(s)} h$ system 
by means of the time-like form factor. 

We calculated in total $46$ channels for possible intermediate $D^\ast_{(s)}$ contributions in $B^0, B^+$ and $B_s^0$ decays, 
and clarified the strong decays $D^\ast_{(s)} \to D_{(s)} h$ by $u$-, $d$- and $s$-quark pair configurations for each resonant structure. 
Concerning on the charged resonance $D^{\ast \pm}$, 
the strong decays with the $u$- and $d$-quark configurations happen by the pole mass dynamics, 
while the decay with $s$-quark configuration happens by the BWT effect. 
For the neutral resonances $\bar D^{\ast 0}$ and $D^{\ast 0}$, 
the strong decays with the $u$-quark configuration happens by the pole mass dynamics, 
and the $d$- and $s$-quark configurations happen by the BWT effect. 
The strong decays of $D_s^\ast$ following from the $B_s$ weak decay happen only by the BWT effect 
with both the $u$- and $d$-quark configurations. 
Our predictions certify the smallness ($ < 5 \%$) of the BWT effect in the three-body $B_{(s)}$ decays with the intermediate resonant states $D^{\ast}$, 
the litter tension between our predictions and the current data of the channels 
$B^+\to \bar D^{\ast 0}\pi^+(K^+) \to D^-\pi^+\pi^+(K^+)$ requires the future measurements with high accuracy. 
For the quasi-two-body decay in channel $B_s^0 \to D_s^{\ast -} \pi^+ \to \bar D^0 K^- \pi^+$ happened by the BWT effect, 
the pQCD prediction is consistent with the current LHCb measurement with in the large error. 
We also checked the narrow width approximation of these resonant states 
by extracting the branching ratios of relative two-body decays from these quasi-two-body processes, 
and found it works well for the CKM-favoured channels.

\begin{acknowledgments}

This work is partly supported by the National Science Foundation of China (NSFC) under the Nos. 11805060, 11975112, 11547038 
and the Joint Large Scale Scientific Facility Funds of the NSFC and CAS under Contract No. U1932110.

\end{acknowledgments}

\appendix 

\section{Decay amplitudes}\label{Amplitudes}

The amplitudes of two-body decays $B_{(s)} \to D^{\ast}_{(s)} \pi, \, D^{\ast}_{(s)} K$ in the factorization approaches are \cite{prd78-014018,jpg37-015002},
\beq
&&{\mathcal A}\big(B^0\to D^{\ast -} \pi^+ \big)=\frac{G_F}{\sqrt2}V^\ast_{cb}V_{ud}
\big[\big(\frac{c_1}{3}+c_2\big)F_{T D^\ast} +c_1M_{T D^\ast}+\big(c_1+\frac{c_2}{3}\big)F_{A \pi} +c_2M_{A \pi}\big]\;,
\label{eq:AB01}\\
&&{\mathcal A}\big(B^0\to D^{\ast +} \pi^- \big)=\frac{G_F}{\sqrt2}V^\ast_{ub}V_{cd}
\big[\big(c_1+\frac{c_2}{3}\big)F_{A D^\ast}+c_2M_{A D^\ast}+\big(\frac{c_1}{3}+c_2\big)F_{T \pi} +c_1M_{T \pi}\big]\;,
\label{eq:AB02}\\
&&{\mathcal A}\big(B^0\to D^{\ast -} K^+ \big)=\frac{G_F}{\sqrt2}V^\ast_{cb}V_{us}
\big[\big(\frac{c_1}{3}+c_2\big)F_{T D^\ast} +c_1M_{T D^\ast}\big]\;,
\label{eq:AB03}\\
&&{\mathcal A}\big(B^0\to \bar D^{\ast 0} \pi^0  \big)=\frac{G_F}{2}V^\ast_{cb}V_{ud}
\big[-\big(c_1+\frac{c_2}{3}\big)F_{T \pi} -c_2M^\prime_{T \pi}+\big(c_1+\frac{c_2}{3}\big)F_{A \pi} +c_2M_{A \pi}\big]\;,\\
&&{\mathcal A}\big(B^0\to D^{\ast 0} K^0 \big)=\frac{G_F}{\sqrt2}V^\ast_{ub}V_{cs}
\big[\big(c_1+\frac{c_2}{3}\big)F_{T K} +c_2M_{T K}\big]\;,\\
&&{\mathcal A}\big(B^0\to \bar D^{\ast 0} K^0 \big)=\frac{G_F}{\sqrt2}V^\ast_{cb}V_{us}
\big[\big(c_1+\frac{c_2}{3}\big)F_{T K} +c_2M^\prime_{T K}\big]\;;
\eeq
\beq
&&{\mathcal A}\big(B^+\to D^{\ast +} \pi^0 \big)=\frac{G_F}{2}V^\ast_{ub}V_{cd}
\big[-\big(\frac{c_1}{3}+c_2\big)F_{A D^\ast} -c_1M_{A D^\ast}+\big(\frac{c_1}{3}+c_2\big)F_{T \pi} +c_1M_{T \pi}\big]\;,\\
&&{\mathcal A}\big(B^+\to \bar D^{\ast 0} \pi^+ \big)=\frac{G_F}{\sqrt2}V^\ast_{cb}V_{ud}
\big[\big(\frac{c_1}{3}+c_2\big)F_{T D^\ast}+c_1M_{T D^\ast}+\big(c_1+\frac{c_2}{3}\big)F_{T \pi} +c_2M^\prime_{T \pi}\big]\;,\\
&&{\mathcal A}\big(B^+\to \bar D^{\ast 0} K^+ \big)=\frac{G_F}{\sqrt2}V^\ast_{cb}V_{us}
\big[\big(\frac{c_1}{3}+c_2\big)F_{T D^\ast} +c_1M_{T D^\ast}+\big(c_1+\frac{c_2}{3}\big)F_{T K} +c_2M^\prime_{T K}\big]\;,\\
&&{\mathcal A}\big(B^+\to D^{\ast 0}  K^+ \big)=\frac{G_F}{\sqrt2}V^\ast_{ub}V_{cs}
\big[\big(\frac{c_1}{3}+c_2\big)F_{A D^\ast} +c_1M_{A D^\ast}+\big(c_1+\frac{c_2}{3}\big)F_{T K} +c_2M_{T K}\big]\;;
\eeq
\beq
&&{\mathcal A}\big(B^0_s\to D^{\ast -}_s  \pi^+ \big)=\frac{G_F}{\sqrt2}V^\ast_{cb}V_{ud}
\big[\big(\frac{c_1}{3}+c_2\big)F_{T D^\ast_s} +c_1M_{T D^\ast_s} \big]\;,\\
&&{\mathcal A}\big(B^0_s\to  D^{\ast -}_s K^+ \big)=\frac{G_F}{\sqrt2}V^\ast_{cb}V_{us}
\big[\big(\frac{c_1}{3}+c_2\big)F_{T D^\ast_s} +c_1M_{T D^\ast_s}+\big(c_1+\frac{c_2}{3}\big)F_{A K} +c_2M_{A K}\big]\;,\\
&&{\mathcal A}\big(B^0_s\to D^{\ast -} \pi^+  \big)=\frac{G_F}{\sqrt2}V^\ast_{cb}V_{us}
\big[\big(c_1+\frac{c_2}{3}\big)F_{A \pi} +c_2M_{A \pi}\big]\;,\\
&&{\mathcal A}\big(B^0_s\to D^{\ast +} K^- \big)=\frac{G_F}{\sqrt2}V^\ast_{ub}V_{cd}
\big[\big(\frac{c_1}{3}+c_2\big)F_{T K} +c_1M_{T K}\big]\;,\\
&&{\mathcal A}\big(B^0_s\to \bar D^{\ast 0} \pi^0 \big)=\frac{G_F}{2}V^\ast_{cb}V_{us}
\big[\big(c_1+\frac{c_2}{3}\big)F_{A \pi} +c_2M_{A \pi}\big]\;, \\
&&{\mathcal A}\big(B^0_s\to \bar D^{\ast 0} \bar K^0 \big)=\frac{G_F}{\sqrt2}V^\ast_{cb}V_{ud}
\big[\big(c_1+\frac{c_2}{3}\big)F_{T K} +c_2M^\prime_{T K}\big]\;,
\eeq
where $c_{i=1,2}(\mu)$ is the tree-level Wilson coefficients carrying 
the physics extending in the energy scale regions from $m_W$ to $m_B$. 
The factorizable and non-factorizable scattering amplitudes $F$ and $M$ carry the physics below the $m_B$ energy scale, 
they are represented by means of the hadron matrix elements with certified four fermion effective operators, 
as expressed in Eq. (\ref{eq:eff-hamilton}). The pQCD calculation result in  
\beq
F_{T D_{(s)}^\ast} &=& 8\pi C_F m^4_B f_{\pi(K)}  \int dx_1 dx_R \int b_1 db_1 b_R db_R \phi_B(x_1,b_1) \phi_{D\pi}(x_R,b_R,s) \nonumber\\
&\times&\big\{\big[\sqrt{\zeta}(2x_R-1)-x_R-1 \big] E^{(1)}_{a}(t_a)h_{a}(x_1,x_R,b_1,b_R)
-\left(\zeta+r_c \right) E^{(2)}_{a}(t_b)h_{b}(x_1,x_R,b_1,b_R) \big\}\,, \\
M_{T D_{(s)}^\ast} &=& 32\pi C_F m^4_B/\sqrt{2N_c}  \int dx_1 dx_R dx_3\int b_1 db_1 b_3 db_3\phi_B(x_1,b_1)\phi_{D\pi}(x_R,b_R,s)\phi^A \nonumber\\
&\times& \big\{ \left[\zeta\left(x_R-x_3+1\right)-x_R\sqrt{\zeta}+x_1+x_3-1)\right]E_{b}(t_c)h_{c}(x_1,x_R,x_3,b_1,b_3)\nonumber\\
&+&\left[x_3\left(1-\zeta\right)+x_R\left(1-\sqrt{\zeta}\right)-x_1 \right]E_{b}(t_d)h_{d}(x_1,x_R,x_3,b_1,b_3) \big\}\,, \\
F_{A D_{(s)}^\ast} &=& 8\pi C_F m^4_B f_B \int dx_R dx_3\int b_R db_R b_3 db_3\phi_D(x_R,b_R,s)\nonumber\\
&\times&\big\{[(1-x_R) \phi^A+2r_0x_R \hat{\zeta}\sqrt{\zeta}\phi^P ] E^{(1)}_{c}(t_e)h_{e}(x_R,x_3,b_R,b_3)\nonumber\\
&+&\big[\left(x_3(\zeta-1)-\zeta\right)\phi^A-r_0r_c[\phi^P+\hat{\zeta}(\zeta+1)\phi^T]\big] E^{(2)}_{c}(t_f)h_{f}(x_R,x_3,b_R,b_3)\big\}\,,\\
M_{A D_{(s)}^\ast} &=& 32\pi C_F m^4_B /\sqrt{2N_c} \int dx_1 dx_R dx_3\int b_1 db_1 b_R db_R \phi_B(x_1,b_1)\phi_{D\pi}(x_R,b_R,s) \nonumber\\
&\times&\big\{\big[[\zeta(x_R+1)+x_1-x_3(\zeta-1)]\phi^A+r_0\hat{\zeta}\sqrt{\zeta}[((1-x_3)(1-\zeta)-x_1)(\phi^P+\phi^T)\nonumber\\
&+& x_R(\phi^T-\phi^P)]\big]E_{d}(t_g)h_{g}(x_1,x_R,x_3,b_1,b_R)+\big[(1-x_R)(\zeta-1)\phi^A-
r_0\hat{\zeta}\sqrt{\zeta}[(x_R-1)(\phi^P+\phi^T)\nonumber\\
&+&(x_1+x_3\zeta-\zeta-x_3)(\phi^T-\phi^P)]\big]E_{d}(t_h)h_{h}(x_1,x_R,x_3,b_1,b_R)\big\}\,; \\
F_{T \pi(K)}&=& 8\pi C_F m^4_B F_{D\pi}(s) \int dx_1 dx_3\int b_1 db_1 b_3 db_3 \phi_B(x_1,b_1)\nonumber\\
&\times&\big\{\big[\phi^A[x_3(1-\zeta)+1]-r_0[\phi^P(2x_3-1)+\phi^T\hat{\zeta}(2 x_3(\zeta-1)+\zeta+1)]\big]E^{(1)}_{e}(t_m) \nonumber\\
&\times& h_m(x_1,x_3,b_1,b_3) +\left[2r_0\hat{\zeta}\phi^P(\zeta(1-x_1)-1)-
\zeta x_1\phi^A\right]E^{(2)}_{e}(t_n) \nonumber\\
&\times&h_n(x_1,x_3,b_1,b_3)\big\}\;,
\eeq
\beq
M_{T \pi(K)} &=& 32\pi C_F m^4_B /\sqrt{2N_c} \int dx_1 dx_R dx_3\int b_1 db_1 b_R db_R\phi_B(x_1,b_1)\phi_{D\pi}(x_R,b_R,s) \nonumber\\
&\times&\big\{\big[  (\zeta-1)(1-x_1-x_R)\phi^A-r_0\hat{\zeta}[\zeta(x_R+x_1)(\phi^P+\phi^T)+x_3(1-\zeta)(\phi^P-\phi^T)-2\zeta\phi^T]\big] \nonumber\\
&\times& E_{f}(t_o)h_{o}(x_1,x_R,x_3,b_1,b_R)-\big[[r_c\sqrt{\zeta}+x_3(\zeta-1)-x_R+x_1]\phi^A+r_0\zeta(x_R-x_1)(\phi^T-\phi^P)\nonumber\\
&+&r_0[x_3(1-\zeta)\phi^P-(4r_c\sqrt{\zeta}+x_3\zeta-x_3)\phi^T]\big]
E_{f}(t_p)h_{p}(x_1,x_R,x_3,b_1,b_R)\big\}\;,\\
M^\prime_{T \pi(K)} &=& 32\pi C_F m^4_B /\sqrt{2N_c} \int dx_1 dx_R dx_3\int b_1 db_1 b_R db_R\phi_B(x_1,b_1)\phi_{D\pi}(x_R,b_R,s) \nonumber\\
&\times&\big\{\big[  \phi^A [x_1+x_R-1-\sqrt{\zeta} r_c-\zeta(x_1+x_R-1)]
-r_0 \hat {\zeta}\zeta [\phi^P(x_1+x_R-x_3)+\phi^T(x_1+x_R+x_3-2)] \nonumber\\
&+& r_0 x_3\hat {\zeta}(\phi^T-\phi^P)\big] E_{f}(t_{o^\prime})h_{o^\prime}(x_1,x_R,x_3,b_1,b_R)-\big[[(\zeta-1)x_3+x_1-x_R]\phi^A+
r_0x_3 (\phi^P+\phi^T)\nonumber\\
&+&r_0\zeta \hat {\zeta}(x_1-x_R)(\phi^T-\phi^P)\big]
E_{f}(t_{p^\prime})h_{p^\prime}(x_1,x_R,x_3,b_1,b_R)\big\}\;,\\
F_{AK} &=& 8\pi C_F m^4_B f_B \int dx_R dx_3\int b_R db_R b_3 db_3 \phi_{D\pi}(x_R,b_R,s))\nonumber\\
&\times&\big\{ \big[ [((\zeta-1)x_3+1)\phi^A+r_0 r_c\phi^P]
-r_0r_c\hat {\zeta}(\zeta+1)\phi^T\big]E^{(1)}_{g}(t_s)h_{s}(x_R,x_3,b_R,b_3)\nonumber\\
&+& \left[-x_R\phi^A+2r_0 \hat {\zeta}\sqrt{\zeta}(\zeta+x_R-1)\phi^P\right]
E^{(2)}_{g}(t_t)h_{t}(x_R,x_3,b_R,b_3) \big\},\\
M_{AK} &=& 32\pi C_F m^4_B /\sqrt{2N_c} \int dx_1 dx_R dx_3\int b_1 db_1 b_3 db_3\phi_B(x_1,b_1)\phi_{D\pi}(x_R,b_R,s)  \nonumber\\
&\times& \big\{ \big[ [(1-\zeta)(x_1+x_R)+\zeta]\phi^A
-r_0 \hat {\zeta}\sqrt {\zeta} [(x_3+\zeta+1-x_3 \zeta)\phi^T+(1-\zeta)(x_3-1)\phi^P] \nonumber\\
&+ &r_0 \hat {\zeta} \sqrt {\zeta} (x_R+x_1)(\phi^T-\phi^P) \big] E_{h}(t_u)h_{u}(x_1,x_R,x_3,b_1,b_R) +\big[ [\zeta(1+x_R-x_1-x_3)+x_3-1] \phi^A \nonumber\\
&+&r_0 \sqrt {\zeta} [ (x_3-1)(\phi^P-\phi^T)+ \hat{\zeta}(x_1-x_R)(\phi^P+\phi^T)] \big]E_{h}(t_v)h_{v}(x_1,x_R,x_3,b_1,b_R) \big\},
\eeq
with $\hat {\zeta}=1/(\zeta-1)$.

The hard scale $t_i$ in the pQCD approach to deal with hard scatterings is chosen as the largest virtuality of the internal momentum transition, 
\beq
t_a &=& {\bf Max} \big\{m_B\sqrt{\vert a_1\vert},m_B\sqrt{\vert a_2\vert}, 1/b_R, 1/b_1 \big\},~~~~~~
t_b= {\bf Max} \big\{m_B\sqrt{\vert b_1\vert},m_B\sqrt{\vert b_2\vert}, 1/b_1, 1/b_R \big\};\nonumber\\
t_c &=& {\bf Max} \big\{m_B\sqrt{\vert c_1\vert},m_B\sqrt{\vert c_2\vert}, 1/b_1, 1/b_3 \big\},~~~~~~~~
t_d= {\bf Max} \big\{m_B\sqrt{\vert d_1\vert},m_B\sqrt{\vert d_2\vert}, 1/b_1, 1/b_3 \big\};\nonumber\\
t_e &=& {\bf Max} \big\{m_B\sqrt{\vert e_1\vert},m_B\sqrt{\vert e_2\vert}, 1/b_R, 1/b_3 \big\},~~~~~~~
t_f= {\bf Max} \big\{m_B\sqrt{\vert f_1\vert},m_B\sqrt{\vert f_2\vert}, 1/b_3, 1/b_R \big\};\nonumber\\
t_g &=& {\bf Max} \big\{m_B\sqrt{\vert g_1\vert},m_B\sqrt{\vert g_2\vert}, 1/b_R, 1/b_1 \big\},~~~~~~~
t_h= {\bf Max} \big\{m_B\sqrt{\vert h_1\vert},m_B\sqrt{\vert h_2\vert}, 1/b_R, 1/b_1 \big\};\nonumber\\
t_m &=& {\bf Max} \big\{m_B\sqrt{\vert m_1\vert},m_B\sqrt{\vert m_2\vert}, 1/b_3, 1/b_1 \big\},~~~~~
t_n= {\bf Max} \big\{m_B\sqrt{\vert n_1\vert},m_B\sqrt{\vert n_2\vert}, 1/b_1, 1/b_3 \big\};\nonumber\\
t_o &=& {\bf Max} \big\{m_B\sqrt{\vert o_1\vert},m_B\sqrt{\vert o_2\vert}, 1/b_1, 1/b_R \big\},~~~~~~~
t_p= {\bf Max} \big\{m_B\sqrt{\vert p_1\vert},m_B\sqrt{\vert p_2\vert}, 1/b_1, 1/b_R \big\};\nonumber\\
t_{o^\prime} &=& {\bf Max} \big\{m_B\sqrt{\vert o_1^\prime\vert},m_B\sqrt{\vert o_2^\prime\vert}, 1/b_1, 1/b_R \big\},~~~~~~~
t_{p^\prime}= {\bf Max} \big\{m_B\sqrt{\vert p_1^\prime\vert},m_B\sqrt{\vert p_2^\prime\vert}, 1/b_1, 1/b_R \big\};\nonumber\\
t_s &=& {\bf Max} \big\{m_B\sqrt{\vert s_1\vert},m_B\sqrt{\vert s_2\vert}, 1/b_3, 1/b_R \big\},~~~~~~~
t_t= {\bf Max} \big\{m_B\sqrt{\vert t_1\vert},m_B\sqrt{\vert t_2\vert}, 1/b_R, 1/b_3 \big\};\nonumber\\
t_u &=& {\bf Max} \big\{m_B\sqrt{\vert u_1\vert},m_B\sqrt{\vert u_2\vert}, 1/b_R, 1/b_1 \big\},~~~~~~
t_v= {\bf Max} \big\{m_B\sqrt{\vert v_1\vert},m_B\sqrt{\vert v_2\vert}, 1/b_R, 1/b_1 \big\}.
\eeq
In the above expressions, the non-dimensional kinematical factors are 
\beq
a_1&=&x_R,~~~~~~~~~~~~~~~~~~~~~~~~~~~~~~~\qquad a_2=x_R x_1;\nonumber\\
b_1&=&r_c^2+x_1-\zeta,~~~~~~~~~~~~~~~~\qquad b_2=a_2;\nonumber\\
c_1&=&a_2,~~~~~~~~~~~~~~~~~~~~~~~~~~~~~~~~\qquad c_2=x_R[x_1-(1-\zeta)(1-x_3)];\nonumber\\
d_1&=&a_2,~~~~~~~~~~~~~~~~~~~~~~~~~~~~~~~~\qquad d_2=x_R[x_1-(1-\zeta ) x_3];\nonumber\\
e_1&=&x_R-1,~~~~~~~~~~~~~~~~~~~~~~~~\qquad e_2=(1-x_R)[x_3(\zeta-1)-\zeta];\nonumber\\
f_1&=&r_c^2+x_3(\zeta-1)-\zeta,~~~~\qquad f_2=e_2;\nonumber\\
g_1&=&e_2,~~~~~~~~~~~~~~~~~~~~~~~~~~~~~~~~\qquad g_2=x_R[x_1+(x_3-1)(1-\zeta)]+1;\nonumber\\
h_1&=&e_2,~~~~~~~~~~~~~~~~~~~~~~~~~~~~~~~~\qquad h_2=(1-x_R)(x_1+x_3\zeta-x_3-\zeta);\nonumber\\
m_1&=&(1-\zeta)x_3,~~~~~~~~~~~~~~~~~~~\qquad m_2=(1-\zeta)x_3 x_1;\nonumber\\
n_1&=&(1-\zeta)x_1,~~~~~~~~~~~~~~~~~~~~\qquad n_2= m_2;\nonumber\\
o_1&=&m_2,~~~~~~~~~~~~~~~~~~~~~~~~~~~~~~~\qquad o_2=(1-x_1-x_R)(x_3\zeta-x_3-\zeta);\nonumber\\
p_1&=&m_2,~~~~~~~~~~~~~~~~~~~~~~~~~~~~~~~\qquad p_2=r_c^2+x_3(\zeta-1)(x_R-x_1);\nonumber\\
o^\prime_1&=&(1-\zeta)x_3 x_1,~~~~~~~~~~~~~~~~\qquad o^\prime_2=r_c^2-(x_R+x_1-1)(x_3(\zeta-1)-\zeta);\nonumber\\
p^\prime_1&=&o_1^\prime,~~~~~~~~~~~~~~~~~~~~~~~~~~~~~~~~\qquad p^\prime_2=x_3(\zeta-1)(x_R-x_1);\nonumber\\
s_1&=&r_c^2-x_3\zeta+x_3-1,~~~~~\qquad s_2=x_R(x_3-1)(1-\zeta);\nonumber\\
t_1&=&x_R(\zeta-1),~~~~~~~~~~~~~~~~~~~\qquad t_2=s_2;\nonumber\\
u_1&=&s_2,~~~~~~~~~~~~~~~~~~~~~~~~~~~~~~~~\qquad u_2=(x_R+x_1-1)(\zeta+x_3-x_3\zeta)+1;\nonumber\\
v_1&=&s_2,~~~~~~~~~~~~~~~~~~~~~~~~~~~~~~~~\qquad v_2=(x_3-1)(1-\zeta)(x_R-x_1) \,.
\label{eq:kinematic-fact}
\eeq

The hard functions $h_i$ ($i \in \{a,b,c,d,e,f,g,h,m,n,o,p,o^\prime,p^\prime,s,t,u,v \}$) in scattering amplitudes 
is expressed in terms of transversal distances $b_i$ conjugated to the transversal momentum by fourier transform. 
\beq
&&h_i(x1,x2,(x3),b_1,b_2) = h_{i1}(\beta,b_2)\times h_{i2}(\alpha,b_1,b_2),\nonumber\\
&&h_{i1}(\beta,b_2) = \left\{\begin{array}{ll}
K_0(\sqrt{\beta}b_2), & \quad  \quad \beta >0\\
\frac{i\pi}{2}H_0^{(1)}(\sqrt{-\beta}b_2),& \quad  \quad \beta<0
\end{array} \right.\nonumber\\
&&h_{i2}(\alpha,b_1,b_2) = \left\{\begin{array}{ll}
\theta(b_2-b_1)I_0(\sqrt{\alpha}b_1)K_0(\sqrt{\alpha}b_2)+(b_1\leftrightarrow b_2), & \quad   \alpha >0\\
\frac{i\pi}{2}\theta(b_2-b_1)J_0(\sqrt{-\alpha}b_1)H_0^{(1)}(\sqrt{-\alpha}b_2)+(b_1\leftrightarrow b_2),& \quad   \alpha<0
\end{array} \right.
\eeq
where $J_0$ is the Bessel function, $K_0$ and $I_0$ are modified Bessel functions, $N_0$ is the Neumann function, 
and $H_0$ is the Hankel function of the first kind with relation $H_0^{(1)}(x)=J_0(x)+iN_0(x)$. 
The kinematic factors $\alpha$ and $\beta$ are exactly the certain case of $i_1$ and $i_2$ 
defined in Eq. (\ref{eq:kinematic-fact}), respectively.

The evolution functions in the scattering amplitudes take into account the strong coupling constant 
and also the Sudakov suppressed factors from the resummations of end-point singularity \cite{prd63-074009,prd63-054008}. 
\beq
&&E^{(1)}_a(t) = \alpha_s(t){\rm exp}[-S_B(t)-S_C(t)] S_t(x_R)\;,\nonumber\\
&&E^{(2)}_a(t) = \alpha_s(t){\rm exp}[-S_B(t)-S_C(t)]  S_t(x_1)\;, \\
&&E_b(t) = \alpha_s(t){\rm exp}[-S_B(t)-S_C(t)-S_P(t)]|_{b_R=b_1}\;,\\
&&E^{(1)}_c(t) = \alpha_s(t){\rm exp}[-S_C(t)-S_P(t)]S_t(x_R)\;,\nonumber\\
&&E^{(2)}_c(t) = \alpha_s(t){\rm exp}[-S_C(t)-S_P(t)]  S_t(x_3)\;, \\
&&E_d(t) = \alpha_s(t){\rm exp}[-S_B(t)-S_C(t)-S_P(t)]|_{b_3=b_R}\;,\\
&&E^{(1)}_e(t) = \alpha_s(t){\rm exp}[-S_B(t)-S_P(t)] S_t(x_3)\;,\nonumber\\
&&E^{(2)}_e(t) = \alpha_s(t){\rm exp}[-S_B(t)-S_P(t)]  S_t(x_1)\;, \\
&&E_f(t) = \alpha_s(t){\rm exp}[-S_B(t)-S_C(t)-S_P(t)]|_{b_3=b_1}\;,\\
&&E^{(1)}_g(t) = \alpha_s(t){\rm exp}[-S_C(t)-S_P(t)] S_t(x_3)\;,\nonumber\\
&&E^{(2)}_g(t) = \alpha_s(t){\rm exp}[-S_C(t)-S_P(t)]  S_t(x_R)\;, \\
&&E_h(t) = \alpha_s(t){\rm exp}[-S_B(t)-S_C(t)-S_P(t)]|_{b_3=b_R}\;.
\eeq

\section{The $D_{(s)} h$ form factor under the $D_{(s)}^\ast$ dominant approximation}\label{Dsast-appro}

Time-like $D_{(s)} h$ form factor is defined by the transition matrix element from vacuum to the $D_{(s)} h$ system sandwiched by a weak current,
\beq
\langle D_{(s)} h \vert \bar{c} \gamma_\mu (1-\gamma_5) q \vert 0 \rangle 
= - \left( \bar{p}_{R \mu} - \frac{m_{D_{(s)}}^2 - m_h^2}{s} p_{R \mu} \right)  F_{D_{(s)}h}^V(s) 
- \frac{m_{D_{(s)}}^2 - m_h^2}{s} \, p_{R \mu} \,  F_{D_{(s)}h}^S(s) \,,
\label{eq:Dsh-ff}
\eeq
in which the on shell conditions of final mesons $p_{D_{(s)}}^2 =m_{D_{(s)}}^2$ and $p_h^2 = m_h^2$ are indicated. 
Multiplying both sides by $\bar{p}_{R}^\mu$, it modifies to 
\beq
\bar{p}_R^\mu \, \langle D_{(s)} h \vert \bar{c} \gamma_\mu (1-\gamma_5) q \vert 0 \rangle 
= - \frac{1}{s} \left[ 2 \left( m_{D_{(s)}}^2 + m_h^2 \right) s - s^2 - \left( m_{D_{(s)}}^2 - m_h^2 \right)^2 \right] F_{D_{(s)}h}^V(s) \,.
\label{eq:Dsh-ff-1}
\eeq
We note that the scalar form factor term is neglected here since the involved $D_{(s)} h$ system is in the ${\rm P}$-wave component. 
In the singe resonance $D_{(s)}^\ast$ approximation, the matrix element from vacuum to $D_{(s)} h$ system is detached to 
\beq
\langle D_{(s)} h \vert \bar{c} \gamma_\mu (1-\gamma_5) q \vert 0 \rangle \rightarrow 
\frac{\langle D_{(s)} h \vert D_{(s)}^\ast \rangle \langle D^\ast_{(s)} \vert \bar{c} \gamma_\mu (1-\gamma_5) q \vert 0 \rangle}
{\left[ m^2_{D_{(s)}} - s - im_{D^\ast_{(s)}} \Gamma_{D^\ast_{(s)}}(s) \right]} 
= \frac{\sqrt{s} \, g_{D^\ast_{(s)} D_{(s)} h} \, f_{D^\ast_{(s)}} \, \sum_{\mu,\nu} \epsilon^\ast_\mu \epsilon_\nu \, \bar{p}_{D_{(s)}^\ast}^\nu}
{\left[ m^2_{D_{(s)}} - s - im_{D^\ast_{(s)}} \Gamma_{D^\ast_{(s)}}(s) \right]} \,.
\label{eq:Dsh-ff-2}
\eeq
To obtain this expression, the definitions $\langle D_{(s)} h \vert D_{(s)}^\ast \rangle = g_{D^\ast_{(s)} D_{(s)} h} \, (\bar{p}^\ast_{D_{(s)}} \cdot \epsilon)$ 
and $\langle D^\ast_{(s)} \vert \bar{c} \gamma_\mu (1-\gamma_5) q \vert 0 \rangle = f_{D^\ast_{(s)}} \, \vert p_{D^\ast_{(s)}} \vert \, \epsilon_\mu^\ast$ 
are used. Substituting Eq. (\ref{eq:Dsh-ff-2}) in to Eq. (\ref{eq:Dsh-ff-1}), we obtain the time-like $D_{(s)} h$ form factor in Eq. (\ref{eq:ff-Dh}).

}

\end{document}